\documentclass[
 prx,
 reprint,
 amsmath,
 amssymb,
 aps,
 superscriptaddress,
]{revtex4-2}

\usepackage{graphicx}
\usepackage{import}
\usepackage{siunitx}
\usepackage{upgreek}
\usepackage{dsfont}
\usepackage[pdfpagemode=UseNone,pdfstartview=FitH,colorlinks=true,linkcolor=blue,citecolor=blue,urlcolor=black]{hyperref}
\usepackage[all]{hypcap}

\begin{document}

\title{Low-leakage superconducting-qubit measurement with sub-100-ns total duration}

\author{Peter A. Spring}
\email{peter.spring@riken.jp}
\affiliation{
RIKEN Center for Quantum Computing (RQC), Wako, Saitama 351-0198, Japan
}

\author{Adrian L. Hesse}
\affiliation{
RIKEN Center for Quantum Computing (RQC), Wako, Saitama 351-0198, Japan
}

\author{Shiyu Wang}
\affiliation{
RIKEN Center for Quantum Computing (RQC), Wako, Saitama 351-0198, Japan
}

\author{Shuhei Tamate}
\affiliation{
RIKEN Center for Quantum Computing (RQC), Wako, Saitama 351-0198, Japan
}

\author{Yasunobu Nakamura}%
\affiliation{
RIKEN Center for Quantum Computing (RQC), Wako, Saitama 351-0198, Japan
}
\affiliation{Department of Applied Physics, Graduate School of Engineering, The University of Tokyo, Bunkyo-ku, Tokyo 113-8656, Japan}

\date{\today}

\begin{abstract}
Fast, accurate, and low-leakage qubit measurement is a key requirement for quantum error correction. Here, we demonstrate measurement of a superconducting transmon qubit with a total duration of 97(1) ns, defined as the time from the start of the measurement pulse until the measurement-induced error on a subsequent $\uppi$-pulse operation falls below $10^{-4}$. By combining a large state-averaged resonator decay rate of $\kappa_\mathrm{eff}/2\pi = \SI{30.8}{\mega\hertz}$ with a dispersive shift close to the optimal SNR-per-photon condition, we achieve an assignment error of 0.17(1)\% using a $58$-$\si{\nano\second}$ measurement pulse, with residual readout photons depleting passively in tens of nanoseconds without an active depletion pulse. Using a repeated-measurement sequence together with a leakage-sensitive measurement, we benchmark the measurement-induced state transitions, finding a per-measurement leakage rate of $2.7(2) \times 10^{-5}$, only twice the background rate and two orders of magnitude below the measurement-induced relaxation rate, which dominates the assignment error. Floquet simulations indicate that the multiphoton resonances present at the operating point are weakly coupled and traversed diabatically, without causing leakage. These results demonstrate that a large resonator decay rate, combined with a dispersive shift close to the optimal SNR-per-photon condition, can enable fast, high-fidelity, low-leakage dispersive readout at small qubit–resonator detuning.
\end{abstract}

\maketitle

\section{\label{sec:introduction} Introduction}
\indent Qubit measurement is a central operation in a quantum processor. It conditions feed-forward operations in teleportation- and measurement-based protocols~\cite{gottesman1999demonstrating, skinner2019measurement, google2023measurement, baumer2024efficient}, and extracts stabilizer syndromes in quantum error correction~\cite{fowler2012surface, terhal2015quantum}. In each role, the measurement must be accurate, because subsequent operations depend on the outcome, and fast, because idle qubits decohere during the measurement. However, in superconducting quantum processors, readout remains slower and more error-prone than single- and two-qubit-gate operations~\cite{google2025quantum}.
\\
\indent The dominant method for measuring superconducting qubits maps the qubit state onto a readout resonator through the dispersive interaction~\cite{blais2004cavity, jeffrey_fast_2014, blais_circuit_2021}. A large resonator decay rate, $\kappa/2\pi \gtrsim 10~\mathrm{MHz}$, means the resonator field responds rapidly to the measurement drive, increasing the signal collected early in the pulse and hence the assignment fidelity achieved with short measurement pulses~\cite{walter_rapid_2017, sunada_fast_2022, swiadek_enhancing_2024, spring_fast_2025}. After the pulse, the same rate enables passive photon depletion within tens of nanoseconds, shortening the ring-down time during which high-fidelity qubit operations are disallowed and potentially removing the need for active photon-depletion schemes~\cite{mcclure_rapid_2016, bultink_active_2016, jerger_dispersive_2024, gautier2025optimal, chatterjee_enhanced_2025}.
\\
\indent Measurement duration and fidelity are also constrained by measurement-induced state transitions (MIST)~\cite{sank_measurement-induced_2016, lescanne2019escape, shillito_dynamics_2022, dumas_measurement-induced_2024, kurilovich2025high, connolly2025full, fechant2025offset, dai_characterization_2026, wang2026probing, hazra2026readout, lin2026mitigation}. The signal acquired in a given time grows with the readout photon number, but the resulting effective drive on the qubit can induce multiphoton resonances between computational and non-computational states, causing leakage~\cite{ shillito_dynamics_2022, dumas_measurement-induced_2024, kurilovich2025high, dai_characterization_2026, wang2026probing, hazra2026readout}. This form of leakage is expected to be more pronounced at small qubit--resonator detuning, where the multiphoton resonances are densely distributed as a function of readout drive frequency~\cite{kurilovich2025high}. This has motivated the exploration of dispersive readout in the high-detuning regime, where the sparsity of multiphoton resonances allows readout to be performed at higher photon numbers before causing leakage~\cite{kurilovich2025high, mencia2025raising, dixit2026millimeter}. However, extracting the maximum signal-to-noise ratio (SNR) per photon requires a dispersive shift $\chi$ comparable to the decay rate, $|\chi|\sim\kappa/2$, and a large $\chi$ is most readily achieved at small qubit--resonator detuning~\cite{blais_circuit_2021}. Establishing whether fast, high-fidelity readout is compatible with low leakage in the small detuning regime is therefore of practical importance.
\\
\indent In this work, we demonstrate fast, high-fidelity, low-leakage measurement of a superconducting qubit operating in the small qubit--resonator detuning regime, with frequency ratio $\omega_\mathrm{r}/\omega_\mathrm{q}=1.24$. By combining a large state-averaged effective resonator decay rate of $\kappa_\mathrm{eff}/2\pi = \SI{30.8}{\mega\hertz}$ with a dispersive shift close to the optimal SNR per photon condition, we achieve an assignment error of 0.17(1)\% with a 58-ns measurement pulse, and we show that qubit operations can resume 97(1) ns after the start of the measurement without active photon depletion. We then benchmark the MIST caused by the two-state measurement using a high-fidelity leakage-sensitive measurement, finding a state-averaged leakage probability of $2.7(2)\times 10^{-5}$ per measurement, which is only twice the background rate and two orders of magnitude below the measurement-induced relaxation rate. Floquet simulations show that, although there are several multiphoton resonances traversed by the measurement, their couplings are weak at the readout photon numbers induced by the two-state measurement. Therefore, they are crossed diabatically, without causing leakage. These results establish that high-fidelity, low-leakage dispersive readout with sub-100-ns total duration is achievable in the small qubit–resonator detuning regime.
\section{\label{sec:section2}Model}
\begin{figure}
\includegraphics[width=1\linewidth]{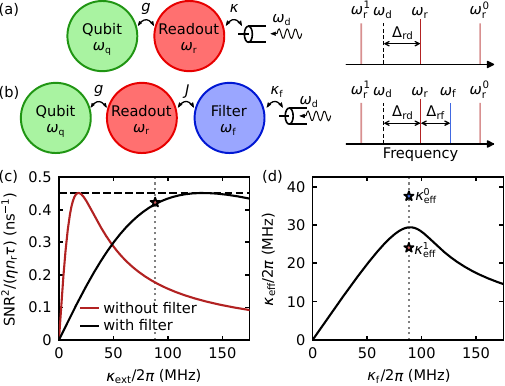}
\caption{\label{fig:optimal_readout_conditions} Optimal readout conditions. (a),(b)~Coupled-mode representations of dispersive readout (a)~without and
(b)~with a dedicated filter resonator. The readout drive and filter resonator are detuned from the mean readout-resonator frequency
$\omega_\mathrm{r}=\left(\omega_\mathrm{r}^0 + \omega_\mathrm{r}^1\right)/2$ by $\Delta_\mathrm{rd} = \omega_\mathrm{r} - \omega_\mathrm{d}$ and $\Delta_\mathrm{rf} = \omega_\mathrm{r} - \omega_\mathrm{f}$, respectively.
(c)~Steady-state $\mathrm{SNR}^2$ per photon per nanosecond, with (black) and without (red) the filter resonator, for $\Delta_\mathrm{rd}=0$ and $\Delta_\mathrm{rf}=0$. The term $\kappa_\mathrm{ext}$ corresponds to $\kappa$ and $\kappa_\mathrm{f}$ for the circuit without and with the filter resonator, respectively. The black dashed line indicates the maximum $\mathrm{SNR}_\mathrm{max}^2/\left(\eta n_\mathrm{r} \tau\right) =8 |\chi|$. (d)~Effective readout-resonator decay rate with a filter resonator, for $\Delta_\mathrm{rf}=0$. In~(c) and~(d), the star markers on the vertical gray dotted lines indicate the expected values for the qubit characterized in this work.}
\end{figure}

We focus on a single transmon qubit from a 64-qubit processor with a tileable 3D-integrated circuit architecture~\cite{tamate2022toward, spring_fast_2025} (see Appendix~\ref{appendix:device_and_setup} for details), featuring dedicated filter resonators, which suppress qubit relaxation through the readout feedline while permitting a large readout bandwidth~\cite{sete_quantum_2015, heinsoo_rapid_2018, swiadek_enhancing_2024, spring_fast_2025}. Before proceeding to the experimental results, we compare the optimal readout parameters with and without a dedicated filter resonator. 
\\
\indent Figure~\ref{fig:optimal_readout_conditions}(a) shows the coupled modes picture of reflection-type dispersive readout without a filter resonator. A qubit with frequency $\omega_\mathrm{q}$ is coupled with coupling strength $g$ to a readout resonator, which is itself coupled to a readout feedline with external coupling $\kappa$. The readout-resonator frequency is qubit-state dependent, and takes values $\omega_\mathrm{r}^0$ and $\omega_\mathrm{r}^1$ for the qubit in $|0\rangle$ and $|1\rangle$, respectively. These frequencies are separated by the dispersive shift $\chi$ through $\omega_\mathrm{r}^1 = \omega_\mathrm{r}^0 + 2\chi$, and we define the mean resonator frequency $\omega_r\equiv \left( \omega_\mathrm{r}^0 + \omega_\mathrm{r}^1\right) /2$. A traveling input field $s_\mathrm{in}$ at drive frequency $\omega_\mathrm{d}$ is applied to the readout feedline, interacts with the readout resonator, and is reflected as the output field $s_\mathrm{out}$ back into the feedline. The SNR of the output field signal acquired over a duration $\tau$ can be expressed as~\cite{bultink_general_2018}
\begin{equation}
    \mathrm{SNR}^2 = 2\eta \int_0^{\tau} \left|s_\mathrm{out}^1(t) - s_\mathrm{out}^0(t) \right|^2 \mathrm{d}t \mathrm{,}
    \label{eq:SNR_definition}
\end{equation}
where $\eta$ is the quantum efficiency of the readout chain and $s_\mathrm{out}^{0}(t)$ and $s_\mathrm{out}^{1}(t)$ are the time-dependent output fields with the qubit in $|0\rangle$ and $|1\rangle$, respectively. We define the drive detuning $\Delta_\mathrm{rd} \equiv \omega_\mathrm{r} - \omega_\mathrm{d}$. Under the condition $\Delta_\mathrm{rd}=0$, the resonator photon number $n_\mathrm{r}$ is qubit-state independent, and in the steady state the maximum $\mathrm{SNR}^2$ per photon, $\mathrm{SNR}_\mathrm{max}^2/n_\mathrm{r} =8\eta |\chi|\tau$, is achieved under the condition $\kappa = 2|\chi|$~\cite{gambetta2008quantum, blais_circuit_2021}.
\\
\indent Figure~\ref{fig:optimal_readout_conditions}(b) shows the coupled modes picture of dispersive readout featuring a dedicated filter resonator~\cite{heinsoo_rapid_2018, swiadek_enhancing_2024}. The readout resonator couples with coupling strength $J$ to a filter resonator with frequency $\omega_\mathrm{f}$, which in turn couples to the readout feedline with external coupling $\kappa_\mathrm{f}$. We define the readout--filter detuning $\Delta_\mathrm{rf} \equiv \omega_\mathrm{r} - \omega_\mathrm{f}$. Under the conditions that $\Delta_\mathrm{rf}=0$ and $\Delta_\mathrm{rd}=0$, the photon number $n_\mathrm{r}$ is again qubit-state independent, and the maximum $\mathrm{SNR}^2$ per photon is once more $\mathrm{SNR}_\mathrm{max}^2/n_\mathrm{r} =8\eta |\chi|\tau$, in this case achieved under the condition $\kappa_\mathrm{f} = 2J^2/|\chi|$. Details of the input--output model used to derive this result are given in Appendix~\ref{appendix:input_output}.
\\
\indent Figure~\ref{fig:optimal_readout_conditions}(c) shows the steady-state $\mathrm{SNR}^2$ per photon per nanosecond for the circuits with and without the filter resonator. The detunings $\Delta_\mathrm{rd}$ and $\Delta_\mathrm{rf}$ are both set to zero. The dispersive shift and readout--filter coupling are $\chi/2\pi = \SI{-9.0}{\mega\hertz}$ and $J/2\pi = \SI{24.4}{\mega\hertz}$, respectively, chosen to match the fitted parameters for the qubit characterized in this work. Without the filter (red curve), the $\mathrm{SNR}$ peaks at $\kappa = 2|\chi|= 2 \pi \times \SI{18}{\mega\hertz}$. With the filter (black curve), the SNR peak instead occurs at $\kappa_\mathrm{p}=2J^2/|\chi| = 2\pi \times \SI{132}{\mega\hertz}$, a large increase of factor $\left(J/\chi\right)^2=7.3$.
\\
\indent For the circuit with the filter resonator, the readout photon number $n_\mathrm{r}$ decays in the long-time limit at the rate $\kappa_{\mathrm{eff}}$, set by the slowest-decaying hybridized readout--filter mode~\cite{heinsoo_rapid_2018, swiadek_enhancing_2024} (see Appendix~\ref{appendix:input_output} for details). The decay rate $\kappa_{\mathrm{eff}}$ is generally state dependent, taking values $\kappa_{\mathrm{eff}}^0$ and $\kappa_{\mathrm{eff}}^1$ for the qubit in $|0\rangle$ and $|1\rangle$, respectively. The two rates are equal only for $\Delta_\mathrm{rf}=0$, where the readout frequencies $\omega_\mathrm{r}^0$ and $\omega_\mathrm{r}^1$ sit symmetrically about the filter frequency. Figure~\ref{fig:optimal_readout_conditions}(d) shows $\kappa_{\mathrm{eff}}$ as a function of $\kappa_\mathrm{f}$ for $\Delta_\mathrm{rf}=0$. The maximum $\kappa_\mathrm{eff}$ value occurs slightly below the critical damping condition, $\kappa_\mathrm{f} = 4J$, which motivates operating close to this point. Since the $\mathrm{SNR}^2$ per photon is maximized under the condition $\kappa_\mathrm{f} = 2J^2/|\chi|$, the critical damping point simultaneously maximizes the $\mathrm{SNR}^2$ per photon when $J=2|\chi|$. Together, these conditions guide the design targets for the readout parameters in this work: $\Delta_\mathrm{rf}=0$, $|\chi|/2\pi=\SI{10}{\mega\hertz}$, $J/2\pi = \SI{25}{\mega\hertz}$, and $\kappa_\mathrm{f}/2\pi=\SI{80}{\mega\hertz}$. Unlike the drive detuning $\Delta_\mathrm{rd}$, the readout--filter detuning $\Delta_\mathrm{rf}$ cannot be tuned \textit{in situ} to enforce $\Delta_\mathrm{rf} = 0$, and in practice we observe a spread of approximately $\pm \SI{10}{\mega\hertz}$. The star markers in Fig.~\ref{fig:optimal_readout_conditions}(c),(d) show the expected $\mathrm{SNR}^2$ per photon and $\kappa_\mathrm{eff}$ for $\kappa_\mathrm{f}/2\pi = \SI{88}{\mega\hertz}$ and $\Delta_\mathrm{rf}/2\pi = \SI{-6}{\mega\hertz}$, matching the fitted values for the qubit characterized in this work (Table~\ref{tab:qubit_readout_parameters}). The predicted $\mathrm{SNR}^2$ per photon per nanosecond is \SI{0.42}{\per\nano\second}, which is \SI{94}{\percent} of the theoretical maximum $\mathrm{SNR}_\mathrm{max}^2/\left(\eta \bar{n}_\mathrm{r} \tau\right) = 8|\chi|$, indicating that the readout is robust to the residual readout--filter detuning. The state-dependent effective decay rates are $\kappa_\mathrm{eff}^0/2\pi = \SI{37.5}{\mega\hertz}$ and $\kappa_\mathrm{eff}^1/2\pi = \SI{24.1}{\mega\hertz}$, corresponding to decay time constants $1/\kappa_\mathrm{eff}^0 = \SI{4.2}{\nano\second}$ and $1/\kappa_\mathrm{eff}^1 = \SI{6.6}{\nano\second}$. These fast decay time constants imply that the readout resonator should respond rapidly to a readout drive, enabling fast qubit measurement.
\section{\label{sec:section3} Readout characterization}
\begin{figure}
\includegraphics[width=1\linewidth]{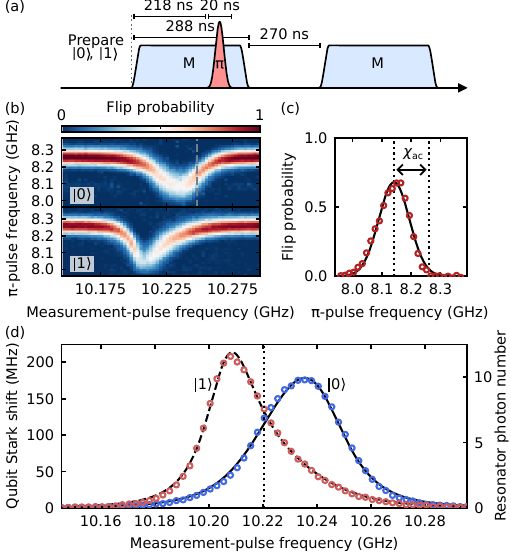}
\caption{\label{fig:parameter_characterization} Readout parameter characterization. (a) Pulse sequence. The carrier frequencies of the first measurement pulse and the qubit $\uppi$ pulse are both swept. (b) Qubit state-flip probability, for the $|0\rangle$- and $|1\rangle$-state preparations. (c) Line cut along the dashed gray line in (b), fit to a normal distribution (black curve). The peak of the fitted distribution gives the ac-Stark-shifted qubit frequency. (d) Magnitude of the ac Stark shift for the $|0\rangle$ and $|1\rangle$ states. The solid(dashed) line shows the fit to the input--output model for the $|0\rangle$$(|1\rangle)$ state. The right axis shows the readout-resonator photon number inferred using $\chi_\mathrm{ac}= 2\chi n_\mathrm{r}$. The dotted vertical line indicates the drive frequency used for the two-state measurement.}
\end{figure}

To characterize the readout parameters, we use the ``chi-kappa-power" method~\cite{sank_system_2025}. Figure~\ref{fig:parameter_characterization}(a) shows the pulse sequence. With the qubit prepared in $|0\rangle$ or $|1\rangle$, a measurement pulse populates the readout resonator and Stark-shifts the qubit, while a concurrent probe pulse drives the qubit. We use a 20-$\si{\nano\second}$ $\uppi$ pulse as the probe. Throughout this work, $\uppi$ pulses use a sum-of-cosine Blackman envelope with DRAG correction~\cite{motzoi2009simple}. A subsequent measurement pulse then reads out the qubit state. We map the probability that the qubit state flips against the frequencies of the first measurement pulse and the qubit probe pulse [Fig.~\ref{fig:parameter_characterization}(b)]. For each readout drive frequency, we fit the flip probability to a normal distribution, and associate the peak to the Stark-shifted qubit frequency~[Fig.~\ref{fig:parameter_characterization}(c)]. We note that the flip probability exhibits a mildly skewed distribution versus the $\uppi$-pulse carrier frequency. In Appendix~\ref{appendix:quantum_model}, we confirm that this skew is reproduced in a simulation of the experiment. The extracted state-dependent Stark shifts are shown in Fig.~\ref{fig:parameter_characterization}(d). The Stark shifts are fit to the expression (see appendix~\ref{appendix:input_output} for details)
\begin{equation}
    \chi_{\mathrm{ac}}^n = 2\chi \left|\frac{J\mathcal{E}}{J^2 - \left(\Delta_\mathrm{fd} - i\frac{\kappa_\mathrm{f}}{2}\right)\Delta_\mathrm{rd}^n}\right|^2 \, \mathrm{,} 
    \label{eq:qubit_stark_shift_vs_drive_frequency}
\end{equation}
where $\mathcal{E}$ is the drive amplitude, $\Delta_\mathrm{fd}\equiv\omega_\mathrm{f}-\omega_\mathrm{d}$ is the detuning of the filter resonator from the drive, $\Delta_\mathrm{rd}^n\equiv \omega_\mathrm{r}^n - \omega_\mathrm{d}$ is the detuning of the state-dependent resonator frequency from the drive, and $n=0,1$ is the qubit state index. The non-Lorentzian response to the readout drive frequency arises due to the dedicated filter resonator. The fitted responses for the $|0\rangle$ and $|1\rangle$ states are shown by the solid and dashed curves, respectively, from which we determine $\omega_\mathrm{r}$, $\omega_\mathrm{f}$, $J$, $\kappa_\mathrm{f}$, and $\chi$. The fitted parameter values are given in Table~\ref{tab:qubit_readout_parameters}.
\begin{table}[b]
\caption{\label{tab:qubit_readout_parameters}
Qubit and readout parameters. Terms $\omega_\mathrm{q}$ and $\alpha$ are the qubit frequency and anharmonicity, respectively.}
\begin{ruledtabular}
\begin{tabular}{cccccccc}
$\omega_\mathrm{q}/2\pi$&
$\alpha/2\pi$&
$\omega_\mathrm{r}^0/2\pi$&
$\omega_\mathrm{f}/2\pi$&
$J/2\pi$&
$\kappa_\mathrm{f}/2\pi$&
$\chi/2\pi$\\
\textrm{(MHz)}&
\textrm{(MHz)}&
\textrm{(MHz)}&
\textrm{(MHz)}&
\textrm{(MHz)}&
\textrm{(MHz)}&
\textrm{(MHz)}\\
\colrule
$8259.7$ & $-360.2$ & $10233$ & $10230$ & $24.4$ & $88$ & $-9.0$ \\
\end{tabular}
\end{ruledtabular}
\end{table}
\section{\label{sec:section4}Fast two-state measurement}
\begin{figure}
\includegraphics[width=1\linewidth]{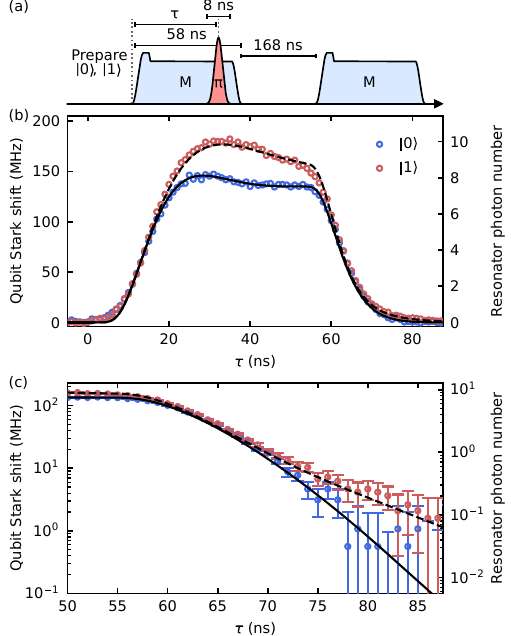}
\caption{\label{fig:resonator_time_response} Readout-resonator time response. (a)~Pulse sequence. The elapsed time $\tau$ between the start of the first measurement and the peak of the $\uppi$ pulse is swept. The $\uppi$-pulse frequency is also swept. (b) Magnitude of the dynamical ac Stark shift for the $|0\rangle$ and $|1\rangle$ state. The solid(dashed) lines show the predictions of the input--output model for the $|0\rangle$($|1\rangle$) state. (c)~Log-scale detail of the ringdown dynamics.}
\end{figure}
We tune up a fast measurement pulse optimized for $|0\rangle$--$|1\rangle$ state discrimination. The 58-\si{\nano\second} duration pulse consists of an initial 9-\si{\nano\second} section with amplitude 1.2 times higher than that of the subsequent flat plateau, to increase the SNR at early times. The rising and falling edges are 6-\si{\nano\second} cosine ramps, included in the total duration. The drive frequency, $\omega_\mathrm{d}/2\pi=\SI{10220.5}{\mega\hertz}$, is close to the frequency that results in a qubit-state-independent resonator photon number in the steady state, as shown by the black dotted line in Fig.~\ref{fig:parameter_characterization}(d). To characterize the dynamics of the readout photon number under this pulse, we use the same approach as that in Refs.~\citenum{sank_system_2025} and~\citenum{beaulieu_fast_2026}. Figure~\ref{fig:resonator_time_response}(a) shows the pulse sequence. With the qubit prepared in $|0\rangle$ or $|1\rangle$, a measurement pulse is applied at the readout frequency, and a frequency-swept probe pulse is applied to the qubit at a swept time $\tau$. We use a short 8-ns $\uppi$ pulse as the probe to improve the time resolution at which the dynamical Stark shift is sampled. A second measurement pulse then reads out the resulting state. The probe pulse frequency that maximizes the flip probability yields the Stark-shifted qubit frequency, which we again determine by fitting to a normal distribution. The extracted Stark shifts for the qubit prepared in $|0\rangle$ and $|1\rangle$ are shown in Fig.~\ref{fig:resonator_time_response}(b). The state-dependent photon-number dynamics predicted by the input--output model is overlaid. The readout drive amplitude $\xi$ and a qubit-readout pulse timing offset are the only fit parameters, with the remaining parameters predetermined from the chi-kappa-power experiment. The model shows good correspondence with the measured photon-number dynamics and reproduces the state-dependent response. The maximum readout photon numbers for the $|0\rangle$ and $|1\rangle$ states are $n_\mathrm{r}^0=8.2$ and $n_\mathrm{r}^1=10.1$, respectively. After the measurement pulse, the photon number decays exponentially at a state-dependent rate [Fig.~\ref{fig:resonator_time_response}(c)], consistent with the input--output model, which predicts decay rates $\kappa_{\textrm{eff}}^0/2\pi=\SI{37.5}{\mega\hertz}$ and $\kappa_{\textrm{eff}}^1/2\pi=\SI{24.1}{\mega\hertz}$ at sufficiently long times after the drive is switched off (see Appendix~\ref{appendix:input_output} for details).
\begin{figure*}
\includegraphics[width=1\linewidth]{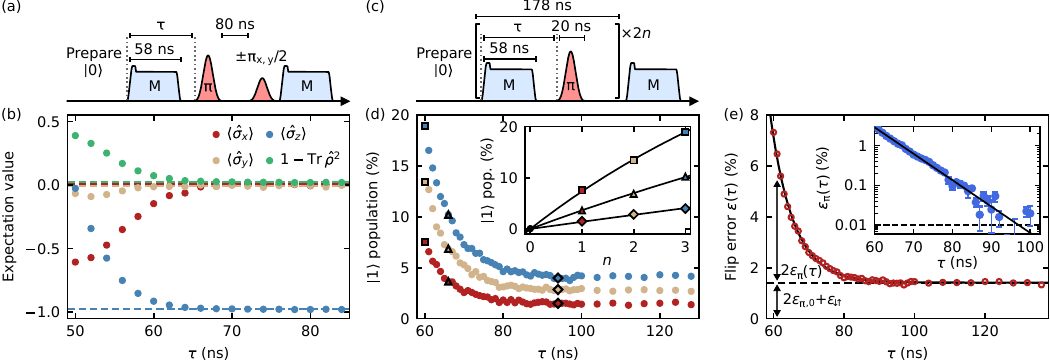}
\caption{\label{fig:total_measurement_duration} Total measurement duration. (a)  Pulse sequence used to perform state tomography of a 20-\si{\nano\second} $\uppi$ pulse applied at time $\tau$ after the start of a measurement pulse. (b) Measured tomography components against the $\uppi$-pulse start time. The dashed lines show the result of a reference tomography experiment with the measurement preceding the $\uppi$ pulse removed. (c) Pulse sequence used to determine the total measurement duration. The time $\tau$ is the interval between the measurement start and the subsequent $\uppi$-pulse start. The measurement--$\uppi$-pulse block is repeated $2n$ times. (d) Excited-state population determined by the final measurement pulse, for $n=1$ (red), $n=2$ (tan), and $n=3$ (blue). The inset shows the excited-state population for three fixed values of $\tau$. The corresponding data points are indicated with black open markers in the main panel. In the inset, the black circle at $n=0$ is the independently-measured SPAM error $\epsilon_\textrm{SPAM}$, and the black curves show the fit to Eq.~\eqref{eq:added_pi_error_model}. (e) Flip error $\epsilon(\tau)$ extracted from (d). The black curve shows the fit to an exponential decay with a constant offset. The fitted offset is shown by the black dashed line. The inset shows the inferred $\uppi$-pulse error contribution from residual readout photons~(see main text).}
\end{figure*}
\\
\indent The rapid readout-photon decay suggests that high-fidelity qubit operations can resume shortly after the measurement pulse, without requiring an active depletion pulse. We test this using the pulse sequence shown in Fig.~\ref{fig:total_measurement_duration}(a). The qubit is first prepared in $|0\rangle$, after which a $58$-\si{\nano\second} measurement pulse is applied at time $\tau=0$, followed by a $\uppi$ pulse at a swept time $\tau$. We then reconstruct the final-state density matrix $\rho$ by quantum state tomography. Figure~\ref{fig:total_measurement_duration}(b) shows the reconstructed final state and its purity as a function of the $\uppi$-pulse start time $\tau$. The residual readout photons due to the measurement pulse induce a qubit Stark shift that produces a non-zero $\sigma_x$ component in the final state. They also dephase the qubit, reducing the final-state purity $\mathrm{Tr}(\hat{\rho}^2)$. The induced Stark shift and the dephasing are both proportional to the readout photon number $n_\mathrm{r}$~\cite{gambetta_qubit-photon_2006}. Due to the large decay rate $\kappa_{\mathrm{eff}}^0$, the reconstructed final state rapidly becomes indistinguishable from that determined in a reference tomography experiment where the measurement before the $\uppi$ pulse is removed (dashed lines).
\\
\indent To determine how long it is necessary to wait after measuring before performing a qubit operation, we define the total measurement duration $t_\mathrm{meas}$ as the time from the start of the measurement pulse until the measurement-induced error on a subsequent $\uppi$ pulse falls below $10^{-4}$. To characterize the total measurement duration, we use the pulse sequence shown in Fig.~\ref{fig:total_measurement_duration}(c). The qubit is initially prepared in $|0\rangle$, followed by a measurement pulse at time $\tau=0$, and a $\uppi$ pulse at swept-time $\tau$. The measurement--$\uppi$-pulse block is then repeated $2n$ times. The duration of each block ($\SI{178}{\nano\second}$) is chosen to ensure the readout photon number is negligible at the start of the subsequent block. A final measurement pulse is then applied, from which the qubit state is assigned by single-shot readout. In the absence of qubit relaxation and $\uppi$-pulse error, the final qubit state will be $|0\rangle$, since the qubit is initialized in $|0\rangle$ and an even number of $\uppi$ pulses are applied prior to the final measurement. In the presence of qubit relaxation and $\uppi$-pulse error, we derive the following expression for the final $|1\rangle$-state population $P_1$~(see Appendix~\ref{appendix:total_measurement_duration} for details)
\begin{equation}
P_1 = \{1 - \left[1-2\epsilon(\tau)\right]^n \}/2 + \epsilon_{\mathrm{SPAM}} \mathrm{,} \label{eq:added_pi_error_model} \\
\end{equation}
where the flip error $\epsilon(\tau)$ is the probability that the qubit transitions from $|0\rangle$ to $|1\rangle$ after the application of two measurement--$\uppi$-pulse blocks and $\epsilon_\textrm{SPAM}$ is a state-preparation and measurement (SPAM) error. To leading order, the flip error can be divided into the following error terms 
\begin{equation}
\epsilon(\tau) = 2\epsilon_\uppi(\tau) + 2\epsilon_{\uppi,0} + \epsilon_{\uparrow\downarrow}  \label{eq:added_pi_error_model_epsilon} \mathrm{.}
\end{equation}
The $\tau$-dependent error term $\epsilon_\uppi\!\left(\tau\right)$ is the state-averaged $\uppi$-pulse error caused by residual readout photons, and the remaining error terms $\epsilon_{\uppi,0}$ and $\epsilon_{\uparrow\downarrow}$ are a background $\uppi$-pulse error and relaxation error, respectively. Figure~\ref{fig:total_measurement_duration}(d) shows the excited-state population for different values of $n$. 
At each time $\tau$, we fit the excited state population to Eq.~\eqref{eq:added_pi_error_model}, as shown in the inset of Fig.~\ref{fig:total_measurement_duration}(d). The flip error $\epsilon(\tau)$ is the fit parameter, with the value $\epsilon_\textrm{SPAM}=5\times10^{-4}$ predetermined from a qubit measurement performed directly following state preparation. The extracted flip errors are shown in Fig.~\ref{fig:total_measurement_duration}(e). To leading order, the added error $\epsilon_\uppi\!\left(\tau\right)$ scales linearly with the photon number, which decays exponentially. Hence, we fit the flip error to the curve $\epsilon(\tau)=Ae^{-\kappa_{\uppi}\tau} + B$, with the relations $Ae^{-\kappa_{\uppi}\tau} = 2\epsilon_\uppi\!\left(\tau\right)$ and $B=2\epsilon_{\uppi,0} + \epsilon_{\uparrow\downarrow}$, where $B$ collects the $\tau$-independent error terms. The fitted value of $B=0.0141$ is dominated by the relaxation error $\epsilon_{\uparrow\downarrow}$ and corresponds to an effective relaxation time $T_{1}^\textrm{eff}=\SI{178}{\nano\second}/0.0141=\SI{12.6}{\micro\second}$. This is consistent with the measurement-induced relaxation time $T_{1}^\mathrm{meas}=\SI{12.7}{\micro\second}$ independently determined in the subsequent section. The fitted value of $\kappa_{\uppi}/2\pi=\SI{24.3(4)}{\mega\hertz}$ is close to the excited-state decay rate $\kappa_\mathrm{eff}^1/2\pi=\SI{24.1}{\mega\hertz}$ predicted independently from the input--output model. This implies that the total measurement duration is limited by the slower readout-photon decay rate $\kappa_\mathrm{eff}^1$ when the qubit is in $|1\rangle$. From the fitted values of $\kappa_{\uppi}$ and $A$, the inferred error due to resonator photons, $\epsilon_\uppi\!\left(\tau\right)$, passes below $1\times10^{-4}$ at $\tau=97(1)~\si{\nano\second}$ [Fig.~\ref{fig:total_measurement_duration}(e) inset]. In Appendix~\ref{appendix:total_measurement_duration}, we use a similar pulse sequence to verify the $\tau$ calibration, by checking that a $\uppi$ pulse performed prior to measurement, ending at time $\tau=0$, is not degraded by the measurement. Thus, we conclude that the total-measurement duration is $t_\mathrm{meas} = 97(1)~\si{\nano\second}$.
\begin{figure}
\includegraphics[width=1\linewidth]{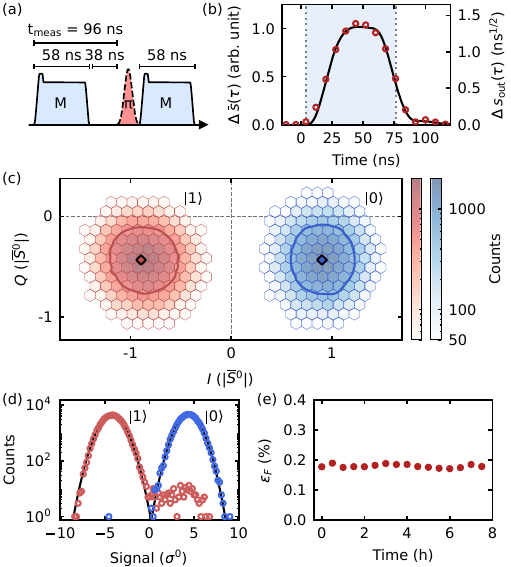}
\caption{\label{fig:two_state_measurement} Two-state measurement. 
(a) Pulse sequence used to determine the assignment fidelity. The sequence is repeated with and without the 20-\si{\nano\second} $\uppi$ pulse. 
(b) Time-dependent signal separation $\Delta \overline{s}(\tau) \equiv \left|\overline{s}^{\,0}(\tau) - \overline{s}^{\,1}(\tau)\right|$ for the postselected second measurement pulse. The $8$-\si{\nano\second} spacing of the data points results from in-FPGA downsampling of the raw samples, which are acquired at $0.5$~GSa/s. The black curve gives the coherent output-field separation $\Delta s_{\textrm{out}}(\tau) \equiv \left|s_{\textrm{out}}^{0}(\tau) - s_{\textrm{out}}^{1}(\tau)\right|$ predicted by the input--output model. The blue-shaded region shows the integration window used for two-state classification.
(c) Hexagonal-bin histograms of the integrated heterodyne signal, with $\num{6e4}$ shots per qubit state. The $I$ and $Q$ signals are normalized by the mean signal magnitude $|\overline{S}^0|$ for the qubit prepared in $|0\rangle$. Bins containing fewer than 50 counts are omitted for clarity. The solid curves show the $68\%$ probability density contours. The diamond markers show the prediction of the input--output model with no adjustable parameters. 
(d) Histograms of the $I$-axis component of the integrated signals. The black curves are fits to normal distributions. The signal is normalized to the fitted standard deviation $\sigma^0$ of the $|0\rangle$-state distribution. (e) Temporal stability of the assignment error $\epsilon_\mathcal{F}$, measured every 30~minutes over a 7.5-hour period.}
\end{figure}
\\
\indent We next characterize the assignment fidelity of the two-state measurement. For single-shot readout, a flux-driven impedance-matched parametric amplifier (IMPA)~\cite{mutus2014strong, roy2015broadband} is used in the phase-preserving mode of operation. The quantum efficiencies are determined to be $\eta=5\%$ and $38\%$ with the IMPA pump turned off and on at the optimal point, respectively, by comparing the measurement-induced dephasing to the measured SNR~\cite{bultink_general_2018}. To characterize the assignment fidelity of the measurement, we use the pulse sequence shown in Fig.~\ref{fig:two_state_measurement}(a). A $58$-\si{\nano\second} measurement pulse to herald the qubit in $|0\rangle$ is followed by a $38$-\si{\nano\second} interval to allow the readout photons to decay, resulting in a $96$-\si{\nano\second} total duration which is close to the value of $t_\mathrm{meas}=\SI{97(1)}{\nano\second}$ determined in the previous experiment. A $20$-\si{\nano\second}-duration idle or $\uppi$ pulse then prepares the qubit in $|0\rangle$ or $|1\rangle$, respectively, immediately followed by a second measurement pulse. We write the single-shot heterodyne record of the second measurement as $s^n(\tau) = s_I^n(\tau) + i\,s_Q^n(\tau)$ for the qubit in state $|n\rangle$ and denote its average as $\overline{s}^{n}(\tau)$. Figure~\ref{fig:two_state_measurement}(b) shows the time dependence of the separation $\Delta \overline{s}(\tau) \equiv \left|\overline{s}^{\,0}(\tau) - \overline{s}^{\,1}(\tau)\right|$. The black curve shows the separation of the output fields $\Delta s_\mathrm{out}(\tau) \equiv \left|s_\mathrm{out}^{\,0}(\tau) - s_\mathrm{out}^{\,1}(\tau)\right|$ predicted by the input--output model. The only fit parameter between the output-field separation $\Delta s_\mathrm{out}(\tau)$ and the measured signal $\Delta \overline{s}(\tau)$ is a magnitude scaling due to the arbitrary units of the measured signal. The blue-shaded region shows the integration window $t_\mathrm{int}$ of 72~\si{\nano\second} used for two-state assignment, chosen to maximize the assignment fidelity. The state separation varies across the integration window, which includes the ring-up and extends beyond the pulse into the ring-down. We therefore apply matched-filter weighting~\cite{gambetta_protocols_2007} to determine the integrated signal $S^n$,
\begin{align}
S^n = \int_0 ^{t_\mathrm{int}} w\!\left(\tau\right) s^n\!\left(\tau\right) \mathrm{d}\tau 
\label{eq:integrated_signal}
\textrm{,} \\
w\!\left(\tau\right) \equiv \left[\overline{s}^{0}\!\left(\tau\right) - \overline{s}^{1}\!\left(\tau\right)\right]^{*} 
\label{eq:matched_filter_weighting_function}
\textrm{.}
\end{align}
Figure~\ref{fig:two_state_measurement}(c) shows a hexagonal-bin histogram of the integrated signal. The signals are normalized by the distance of the mean $|0\rangle$-state signal from the origin, $|\bar{S}^0|$. The diamond markers show the signals $S_\mathrm{out}^0$ and $S_\mathrm{out}^1$ predicted by the input--output model, using the predetermined readout parameters. The same weighting function, integration window, and normalization are applied, which allows comparison to the measured signals without adjustable parameters (see Appendix~\ref{appendix:input_output} for details). The agreement with the data is good, demonstrating that the input--output model can predict the relative integrated signal positions in the IQ plane. Figure~\ref{fig:two_state_measurement}(d) shows the histogram of the $I$-axis component of the integrated signal $S_I$. From normal-distribution fits to the $|0\rangle$- and $|1\rangle$-state distributions, we determine the measurement SNR to be $\textrm{SNR}=8.35$, which implies a low SNR-limited assignment error of $\epsilon_{\textrm{SNR}} = 0.0015\%$~\cite{gambetta_protocols_2007, swiadek_enhancing_2024}. We assign the qubit state from the integrated heterodyne signal using a logistic regression (LR) model. We train the model on a dataset of $10^4$ shots per qubit state. We then test it on ten separate datasets each having $\num{6e4}$ shots per qubit state. We use the assignment fidelity $\mathcal{F}\equiv \left[\mathcal{P}(0|0) + \mathcal{P}(1|1)\right]/2$, where $\mathcal{P}\left(a|b\right)$ is the probability that the second measurement assigns the qubit to state $|a\rangle$, given that it was heralded in $|0\rangle$ and then prepared in state $|b\rangle$. The resulting assignment error $\epsilon_\mathcal{F} \equiv 1-\mathcal{F}$ is $\epsilon_\mathcal{F} = 0.17(1)\%$. In Fig.~\ref{fig:two_state_measurement}(e), we examine the temporal stability of the assignment error over a 7.5-hour period, observing that it remains stable with a mean value $\epsilon_\mathcal{F} = 0.18(1)\%$.
\section{\label{sec:section5}Measurement-induced leakage}
\begin{figure}
\includegraphics[width=1\linewidth]{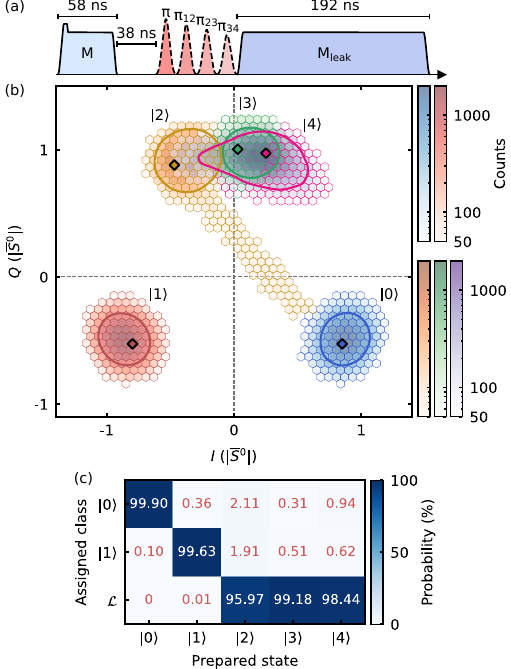}
\caption{\label{fig:Leakage_sensitive_measurement} Leakage-sensitive measurement. (a)~Pulse sequence. The terms $\mathrm{M}$ and $\mathrm{M}_\mathrm{leak}$ denote the two-state and leakage-sensitive measurements. The sequence contains a succession of $20$-\si{\nano\second} $\uppi$ pulses to prepare the qubit in states $|0\rangle$ to $|4\rangle$. (b)~Hexagonal-bin histograms of the weighted $IQ$ signals, with $\num{6e4}$ shots per qubit state. The $I$ and $Q$ signals are normalized by the mean signal magnitude $|\overline{S}^0|$ given the qubit is prepared in the $|0\rangle$ state. Hexagonal bins containing fewer than 50 counts are omitted for clarity. The solid curves show the $68\%$ probability density contours. The diamond markers show the prediction of the input--output model with no adjustable parameters. (c)~Assignment matrix for the leakage sensitive measurement.}
\end{figure}

\indent To benchmark the leakage caused by the two-state-measurement pulse, we first tune up a leakage-sensitive measurement. For single-shot readout, we tune the IMPA to the same optimal point as used for two-state measurement. To characterize the assignment fidelity of the leakage-sensitive measurement, we use the pulse sequence in Fig.~\ref{fig:Leakage_sensitive_measurement}(a). The qubit is heralded in $|0\rangle$ by a two-state measurement, and is then prepared in states $|0\rangle$ to $|4\rangle$ by a 20-\si{\nano\second} idle or the appropriate sequence of 20-\si{\nano\second}-duration $\uppi$ pulses. This is followed by a 192-\si{\nano\second} flat-top leakage-sensitive measurement pulse driven at frequency $\omega_{\mathrm{d},\mathrm{leak}}/2\pi=\SI{10210}{\mega\hertz}$. From the acquired signal of the leakage measurement, we construct the weighted path,
\begin{equation}
    X^n(t) = \int_0 ^{t} w\!\left(\tau \right) s^n(\tau) \mathrm{d}\tau \textrm{,}
    \label{eq:integrated_path} 
\end{equation}
using the matched-filter weighting function defined in Eq.~\eqref{eq:matched_filter_weighting_function}. The conventional integrated signal corresponds to the terminus of this path, $S^n = X^n(t_\mathrm{int})$, with $t_{\mathrm{int}}=\SI{192}{\nano\second}$. Figure~\ref{fig:Leakage_sensitive_measurement}(b) shows the hexagonal-bin histogram of the integrated signals $S^n$ for the different qubit state preparations. Due to the choice of the drive frequency $\omega_{\mathrm{d},\mathrm{leak}}$, the $|2\rangle$-, $|3\rangle$- and $|4\rangle$-state signals form a cluster that is well separated from the $|0\rangle$- and $|1\rangle$-state signals. The diamond markers show the mean signals predicted by the input--output model, with no adjustable parameters, showing good agreement with the measured signals for all the prepared states. Interestingly, the histogram shows a measurement-induced transition $|2\rangle$$\rightarrow$$|0\rangle$. In Appendix~\ref{appendix:measurement_induced_2_0_relaxation}, we analyze the power and frequency dependence of this transition, and find it consistent with a four-wave scattering process that excites a high-frequency spurious mode at \SI{26}{\giga\hertz}~\cite{dai_characterization_2026}.
\\
\indent We perform classification using the signature of the path $X(t)$ as the feature set~\cite{reizenstein2020algorithm, chevyrev2025primer, cao2024superconducting}, rather than the conventional integrated signal $S$. The signature is effective at detecting mid-measurement state transitions such as the observed $|2\rangle$$\rightarrow$$|0\rangle$ transition, improving the assignment fidelity~\cite{cao2024superconducting}. We calculate the depth-5 log-signature of the path with time augmentation~\cite{chevyrev2025primer} and use a multinomial LR model to classify the qubit state to one of the three classes: $|0\rangle$, $|1\rangle$, or $\mathcal{L}$, where $\mathcal{L}$ denotes any leakage state. To reduce false-positive (FP) leakage detection, we additionally apply a thresholding technique. Details of the classifier and thresholding technique are given in Appendix~\ref{appendix:leakage_sensitive_measurement}.
\\
\indent Figure~\ref{fig:Leakage_sensitive_measurement}(c) shows the assignment matrix. We define the leakage-sensitive assignment fidelity as ${\mathcal{F}_\textrm{leak} \equiv \left[\mathcal{P}\left(0|0\right) +\mathcal{P}\left(1|1\right) + \mathcal{P}_N\!\left(\mathcal{L}|\mathcal{L}\right) \right]/3 }$, where $\mathcal{P}_N(\mathcal{L}|\mathcal{L}) \equiv (N-1)^{-1} \sum_{n=2}^{N} \mathcal{P}(\mathcal{L}|n)$ is the probability of assigning the qubit as leaked, averaged over preparation in leakage states $|2\rangle$ to $|N\rangle$. Since we can prepare the qubit in leakage states $|2\rangle$, $|3\rangle$, and $|4\rangle$, we use $\mathcal{P}_4\!\left(\mathcal{L}|\mathcal{L}\right) = \left[\mathcal{P}\left(\mathcal{L}|2\right) + \mathcal{P}\left(\mathcal{L}|3\right) + \mathcal{P}\left(\mathcal{L}|4\right) \right]/3 $. This expression is expected to remain a good approximation for $\mathcal{P}_N\!\left(\mathcal{L}|\mathcal{L}\right)$ with $N > 4$, because for the chosen readout frequency the signals for leakage states above $|4\rangle$ are predicted to lie in close proximity to the $|3\rangle$- and $|4\rangle$-state signals in the IQ plane (See Appendix~\ref{appendix:leakage_sensitive_measurement} for details). The assignment error, $\epsilon_{\mathcal{F},\textrm{leak}} \equiv 1 - \mathcal{F}_\textrm{leak}$, is $\epsilon_{\mathcal{F},\textrm{leak}}=0.87\%$, limited by the $95.97\%$ assignment accuracy when the qubit is prepared in $|2\rangle$, which stems from the measurement-induced $|2\rangle$$\rightarrow$$|0\rangle$ transition. The FP leakage rate, $\mathcal{L}_\textrm{FP} \equiv \left[\mathcal{P}(\mathcal{L}|0) + \mathcal{P}(\mathcal{L}|1) \right]\!/2$, is low, at $\mathcal{L}_\textrm{FP}=0.005\%$. In the subsequent experiment, this low rate enables detection of small leakage populations, which would otherwise be obscured.
\begin{figure}
\includegraphics[width=1\linewidth]{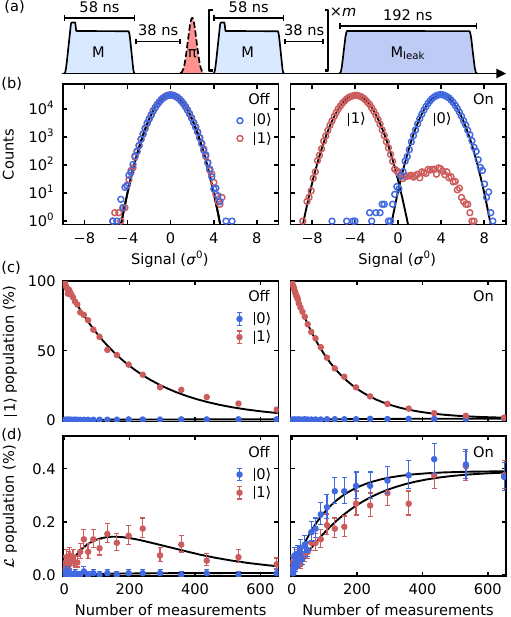}
\caption{\label{fig:MIST_benchmarking} Measurement-induced state transition (MIST) benchmarking. (a) Pulse sequence. The terms $\mathrm{M}$ and $\mathrm{M}_\mathrm{leak}$ denote the two-state and leakage-sensitive measurements. The measurement–ringdown block is repeated $m$ times, with its amplitude set to zero (``off") or to the two-state-assignment optimum (``on").
(b) Projected histograms of the signal acquired from the measurement directly following state preparation, for the off and on sequences. Black curves are fitted normal distributions.
(c),(d) Final $|1\rangle$-class and $\mathcal{L}$-class populations, respectively, from the leakage-sensitive measurement, again shown for the off and on sequences. Black curves are fits to the three-class model with
fixed transition-per-measurement rates.}
\end{figure}
\\
\indent Multiple protocols have been devised to characterize measurement-induced leakage~\cite{hazra_benchmarking_2025, connolly2025full}. Here, we use a sequence that accumulates the leakage error and allows simultaneous characterization of measurement-induced relaxation and assignment fidelity. The pulse sequence is shown in Fig.~\ref{fig:MIST_benchmarking}(a). The qubit is heralded in $|0\rangle$ by a two-state measurement. An idle or $\uppi$ pulse then prepares the qubit in $|0\rangle$ or $|1\rangle$, followed by a two-state measurement which is repeated $m$ times and a final leakage-sensitive measurement. The sequence is repeated with the amplitude of the repeated measurement set to zero, denoted ``off", and set to the optimum for two-state assignment, denoted ``on". For each value of $m$, the sequence is repeated $1.5 \times 10^4$ times to gather statistics. We denote the first of the repeated measurements as the ``assignment measurement". Signal is acquired for three measurements in the sequence: the heralding measurement, the assignment measurement, and the leakage-sensitive measurement. Up to the assignment measurement, the sequence is identical to that used to determine the assignment fidelity, and thus it also serves to benchmark the assignment fidelity. Figure~\ref{fig:MIST_benchmarking}(b) shows histograms of the integrated signal component $S_I$ for the assignment measurement. Since $m$ is swept over 28 values between $m=1$ and $m=650$, the histograms have $28 \times 1.5 \times 10^4 = 4.2\times 10^5$ shots per qubit state. For the off sequence, the $|0\rangle$ and $|1\rangle$ distributions are indistinguishable, as expected. For the on sequence, they become well separated, and assigning the shots using the pretrained two-state LR model results in an assignment error of $\epsilon_\mathcal{F} = 0.17\%$, matching that found in the previous two-state-assignment experiment.
\\
\indent For the leakage-sensitive measurement signal, we calculate the depth-5 log-signature of the path $X(t)$ and use the pretrained leakage-sensitive LR model to classify the final qubit state to one of the three classes $|0\rangle$, $|1\rangle$, or $\mathcal{L}$. We fit the resulting class populations to a three-class model with fixed per-measurement transition probability $P_{ij}$ from class $i$ to class $j$ (see Appendix~\ref{appendix:MIST} for details). Figure~\ref{fig:MIST_benchmarking}(c) shows the final $|1\rangle$-class population for the off and on sequences. The fitted transition probabilities are $P_{10}^{\textrm{off}}=4.73\times10^{-3}$ and $P_{10}^{\textrm{on}}=7.55\times 10^{-3}$, respectively. Taking the total measurement duration $t_{\textrm{meas}}=\SI{96}{\nano\second}$, these correspond to relaxation times $T_1=t_{\textrm{meas}}/P_{10}^{\textrm{off}}=\SI{20.3}{\micro\second}$ and $T_1^{\textrm{meas}}=t_{\textrm{meas}}/P_{10}^{\textrm{on}}=\SI{12.7}{\micro\second}$, demonstrating substantial measurement-induced relaxation~\cite{thorbeck_readout-induced_2024, huang2026readout}. Using the integration-time $t_{\mathrm{int}}=\SI{72}{\nano\second}$ and the value of $T_1^{\textrm{meas}}$, we estimate the coherence-limited assignment error as $\epsilon_{\textrm{cl}}\approx t_{\mathrm{int}}/(4 T_1^{\textrm{meas}})=0.14\%$ \cite{gambetta_protocols_2007, spring_fast_2025}, which accounts for $80 \%$ of the measured two-state assignment error. Thus, we conclude that the assignment error is dominantly caused by $|1\rangle$$\rightarrow$$|0\rangle$ relaxation. Figure~\ref{fig:MIST_benchmarking}(d) shows the final $\mathcal{L}$-class population. For the off sequence, the measured leakage population when the qubit is prepared in $|0\rangle$ is negligible, with fitted probability ${P_{0\mathcal{L}}^{\mathrm{off}}=\num{1(1)e-6}}$. However, there is a detectable $|1\rangle$$\rightarrow$$\mathcal{L}$ transition probability $P_{1\mathcal{L}}^{\textrm{off}}=2.6(5)\times10^{-5}$, attributed to background $|1\rangle$$\rightarrow$$|2\rangle$ heating mechanisms, such as qubit-control-line noise and quasiparticle tunneling at the qubit junction~\cite{chen2016measuring, serniak2018hot, liu2024quasiparticle}. When the measurement is turned on, the leakage probabilities change to $P_{0\mathcal{L}}^{\textrm{on}}=3.6(3)\times10^{-5}$ and $P_{1\mathcal{L}}^{\textrm{on}}=1.8(3)\times10^{-5}$. The measurement significantly raises the leakage rate from $|0\rangle$, while the rate from $|1\rangle$ decreases slightly, which may be a fitting artifact since the decrease is comparable to the fit uncertainty. The mean leakage-per-measurement probabilities, $P_\mathcal{L} \equiv \left(P_{0\mathcal{L}} + P_{1\mathcal{L}}\right)/2$, are $P_\mathcal{L}^{\textrm{off}} = 1.3(3)\times10^{-5}$ and $P_\mathcal{L}^{\textrm{on}}=2.7(2)\times10^{-5}$ for the off and on sequences, respectively. Measurement-induced leakage is thus small and more than two orders of magnitude below the measurement-induced relaxation rate. The fitted $|0\rangle$$\rightarrow$$|1\rangle$ transition probability also increases when the measurement is turned on, from $P_{01}^{\mathrm{off}}=2.3(1)\times10^{-5}$ to $P_{01}^{\mathrm{on}}=6.0(2)\times10^{-5}$. The fitted transition probabilities are summarized in Table~\ref{tab:MIST_transition_probabilities}. In Appendix~\ref{appendix:MIST}, we examine the power dependence of the MIST rates, observing increasing leakage at higher measurement powers.
\begin{table}
\caption{\label{tab:MIST_transition_probabilities}
Per-measurement transition probabilities extracted from the MIST benchmarking sequence, with the repeated-measurement amplitude set to zero (``off'') and to the optimum for two-state assignment (``on'').}
\begin{ruledtabular}
\begin{tabular}{ccccc}
& $P_{10}$ & $P_{01}$ & $P_{0\mathcal{L}}$ & $P_{1\mathcal{L}}$ \\
\colrule
\noalign{\vskip 2pt}
Off & $\num{4.73e-3}$ & $2.3(1)\times10^{-5}$ & $1(1) \times 10^{-6}$ & $2.6(5) \times 10^{-5}$ \\
On  & $\num{7.55e-3}$ & $6.0(2)\times10^{-5}$ & $3.6(3) \times 10^{-5}$ & $1.8(3) \times 10^{-5}$ \\
\end{tabular}
\end{ruledtabular}
\end{table}
\section{\label{sec:section6}Discussion}
\begin{figure}
\includegraphics[width=1\linewidth]{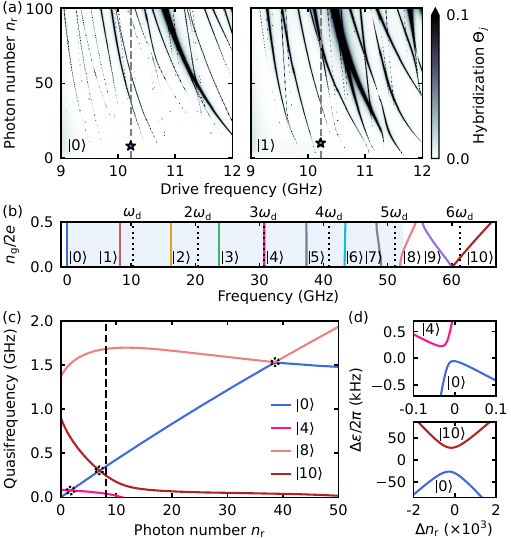}
\caption{\label{fig:multiphoton_resonance_leakage} MIST by multiphoton resonances.
(a) Floquet simulations of the hybridization parameter $\Theta_j$ for the qubit in the $|0\rangle$ and $|1\rangle$ states, at gate charge $n_\mathrm{g}/2e=0.18$. The drive frequency and maximum photon numbers for the two-state measurement in the experiments are indicated by the star markers. (b)~ Energy spectrum of the transmon. The shaded blue
region indicates energies within the $E_J$ potential well. Multiples of the readout drive frequency are indicated by the black dotted lines. (c)~Simulated $|0\rangle$-state quasifrequency at gate charge $n_\mathrm{g}/2e=0.18$. The black dashed line indicates the maximum photon number driven by the two-state measurement. The dashed circles indicate quasifrequency collisions with other transmon states. (d)~Details of the ${|0\rangle}$--${|4\rangle}$ and ${|0\rangle}$--${|10\rangle}$ quasifrequency collisions, revealing weak avoided crossings.}
\end{figure}

Given that the readout is in the small-negative detuning regime, with $\omega_\mathrm{r}/\omega_\mathrm{q}=1.24$, where we expect a dense spectrum of multiphoton resonances versus the readout drive frequency~\cite{shillito_dynamics_2022, dumas_measurement-induced_2024, kurilovich2025high}, the low leakage is perhaps surprising. To determine whether the two-state measurement pulse traverses multiphoton resonances,
we compute the hybridization parameter $ \Theta_j = 1 - |\langle \tilde{j} | \bar{j} \rangle|^2 $, which quantifies the deviation of the Floquet state $|\tilde{j}\rangle$ from the ideal-displaced state $|\bar{j}\rangle$ expected in the absence of multiphoton resonances~\cite{xiao2025diagrammatic, kurilovich2025high,
dai_characterization_2026, floquet2024}. The hybridization parameter $\Theta_j$ vanishes off resonance and reaches a maximum of $1/2$ on resonance. The coherent readout photon number is treated as an effective direct drive on the qubit with driving amplitude $\mathcal{E} = 2g\sqrt{n_r}$~\cite{lledo2023cloaking, dumas_measurement-induced_2024}. Figure~\ref{fig:multiphoton_resonance_leakage}(a) shows the hybridization for different readout-drive frequencies, for the qubit prepared in $|0\rangle$ and $|1\rangle$, computed with the \texttt{Floquet} library~\cite{floquet2024}. The simulations indeed show many multiphoton resonances, visible as the black threadlike features. However, up to the maximum photon number induced by the two-state measurement pulse, $n_{\mathrm{r},\mathrm{max}}^0=8.2$ and $n_{\mathrm{r},\mathrm{max}}^1=10.1$ for the qubit in $|0\rangle$ and $|1\rangle$, respectively, the resonances are very narrow with respect to drive frequency, indicating that the coupling between the qubit levels involved in the resonance is small. In Appendix~\ref{appendix:multiphoton_simulations}, we confirm that this behavior holds across a sweep of gate charge.
\\
\indent To quantify the leakage induced by these weak multiphoton resonances, we focus on the readout drive frequency $\omega_\mathrm{d}/2\pi = \SI{10220.5}{\mega\hertz}$ used for the two-state measurement. Figure~\ref{fig:multiphoton_resonance_leakage}(b) shows the qubit-energy spectrum as a function of the gate charge. Three multiphoton resonances featuring the $|0\rangle$ state are suggested, a $|0\rangle$\nobreakdash--$|4\rangle$ resonance involving three drive photons, a $|0\rangle$\nobreakdash--$|8\rangle$ resonance involving five drive photons, and a $|0\rangle$\nobreakdash--$|10\rangle$ resonance involving six drive photons. Figure~\ref{fig:multiphoton_resonance_leakage}(c) tracks the simulated quasifrequency~\cite{shillito_dynamics_2022, dumas_measurement-induced_2024, floquet2024} for the $|0\rangle$ state for gate charge $n_\mathrm{g}/2e=0.18$. The three suggested resonances all appear as collisions in the quasifrequency spectrum. The $|0\rangle$--$|4\rangle$ and $|0\rangle$--$|10\rangle$ collisions are both traversed by the measurement pulse, which populates the resonator with up to $8.2$ photons (black dashed line). The avoided crossings for these collisions are shown in Fig.~\ref{fig:multiphoton_resonance_leakage}(d). The coupling at the $|0\rangle$--$|4\rangle$ crossing is extremely weak, with splitting $\Delta_\varepsilon/2\pi = \SI{0.3}{\kilo\hertz}$, since this resonance occurs at a readout photon number $n_\mathrm{r}=1.8$ that is too low to drive the transition. The coupling at the $|0\rangle$--$|10\rangle$ crossing is also weak, with splitting $\Delta_\varepsilon/2\pi = \SI{54}{\kilo\hertz}$, since this process involves six drive photons and the $|10\rangle$ state is above the cosine potential barrier, where the transition amplitudes are exponentially suppressed~\cite{kurilovich2025high}. The adiabatic (i.e., leakage-inducing) transition probability is given by the Landau-Zener formula $P_\mathrm{L} = 1-\mathrm{exp}\left(-\pi \Delta_\varepsilon^2 /2\upsilon \right)$~\cite{ikeda2022floquet, dumas_measurement-induced_2024, wang2026probing}, where $\upsilon$ is the effective crossing speed. Due to the large effective decay rate of the readout photons, $\kappa_\mathrm{eff}$, the crossings are traversed rapidly during both the resonator ring-up and ring-down, reducing the chance of leakage at both traversals. For the $|0\rangle$--$|4\rangle$ crossing, the leakage probability is negligible. For the $|0\rangle$--$|10\rangle$ crossing, the leakage probability is $P_\mathrm{L}=\num{1e-7}$, using the input--output model to estimate the crossing speeds $\upsilon_{\mathrm{u}}/2\pi=0.56~\si{\giga\hertz}/\si{\nano\second}$ and $\upsilon_{\mathrm{d}}/2\pi=0.68~\si{\giga\hertz}/\si{\nano\second}$ for the ring-up and ring-down traversals, respectively (see Appendix~\ref{appendix:multiphoton_simulations} for details). In Appendix~\ref{appendix:multiphoton_simulations}, we consider the dominant $|0\rangle$$\rightarrow$$|10\rangle$ leakage probability as a function of the gate charge $n_\mathrm{g}$ and find the maximum leakage rate across 100 evenly sampled gate charges to be $P_\mathrm{L}=\num{1e-6}$. We also perform a similar analysis for the $|1\rangle$ state, finding a similarly low maximum leakage rate of $P_\mathrm{L}=\num{2e-6}$ for the dominant $|1\rangle$$\rightarrow$$|9\rangle$ transition. Therefore, despite operating in the small-detuning regime, the maximum photon number reached during the two-state measurement is insufficient to drive the multiphoton resonance transitions effectively, and the large effective decay rate ensures that the resulting weak resonances are traversed too rapidly to induce leakage. The two-fold increase in leakage during measurement relative to the background may instead originate from inelastic scattering of readout photons into a spurious package mode or a TLS~\cite{connolly2025full, dai_characterization_2026}.
\\
\indent This analysis highlights two advantages of the small-detuning regime. First, whereas the dominant multiphoton resonances at large detuning involve a single drive photon, all resonances here involve at least three drive photons, which suppresses their coupling. Second, a given dispersive shift is achieved with a smaller coupling $g$, so the effective qubit drive $\mathcal{E}=2g\sqrt{n_\mathrm{r}}$, and therefore the transition probability, is weaker at a given photon number.
\section{\label{sec:section7}Conclusions}
\indent We perform dispersive readout on a superconducting qubit using a dedicated filter resonator, combining a large effective decay rate $\kappa_\mathrm{eff}$ with a dispersive shift close to the optimal SNR-per-photon condition to achieve an assignment error of $\epsilon_\mathcal{F} = 0.17(1)\%$ with a $58$-\si{\nano\second} measurement pulse. Rapid passive photon depletion allows high-fidelity qubit operations to resume $97(1)$~\si{\nano\second} after the start of the measurement. Benchmarking with a leakage-sensitive measurement yields a measurement-induced leakage probability $P_\mathcal{L}^{\mathrm{on}} = 2.7(2)\times10^{-5}$, only twice the background probability and consistent with simulations showing that the multiphoton resonances at the readout frequency are weakly coupled and traversed diabatically. These results demonstrate that fast, high-fidelity, low-leakage dispersive readout can be achieved in the small qubit--resonator detuning regime.
\begin{acknowledgments}
We thank the whole Superconducting Quantum Electronics Research Team at RQC for fabrication of the device and assembly and maintenance of the experimental setup. We thank Shoichi Shiba and Aki Dote for assistance with the pattern design, Tsuyoshi Takahashi for performing the Josehpson-junction laser annealing, and Yuji Sakoda for fabrication of the device. We thank Wei Dai for a fruitful discussion of drive-induced state transitions. This work was supported in part by the Ministry of Education, Culture, Sports, Science and Technology (MEXT) Quantum Leap Flagship Program (Q-LEAP) (Grant No. JPMXS0118068682) and the Japan Science and Technology Agency (JST) Adopting Sustainable Partnerships for Innovative Research Ecosystem (ASPIRE) (Grant Number JPMJAP2513).
\end{acknowledgments}

\appendix

\section{\label{appendix:device_and_setup}Device and Experimental Setup}
\begin{figure}
\includegraphics[width=1\linewidth]{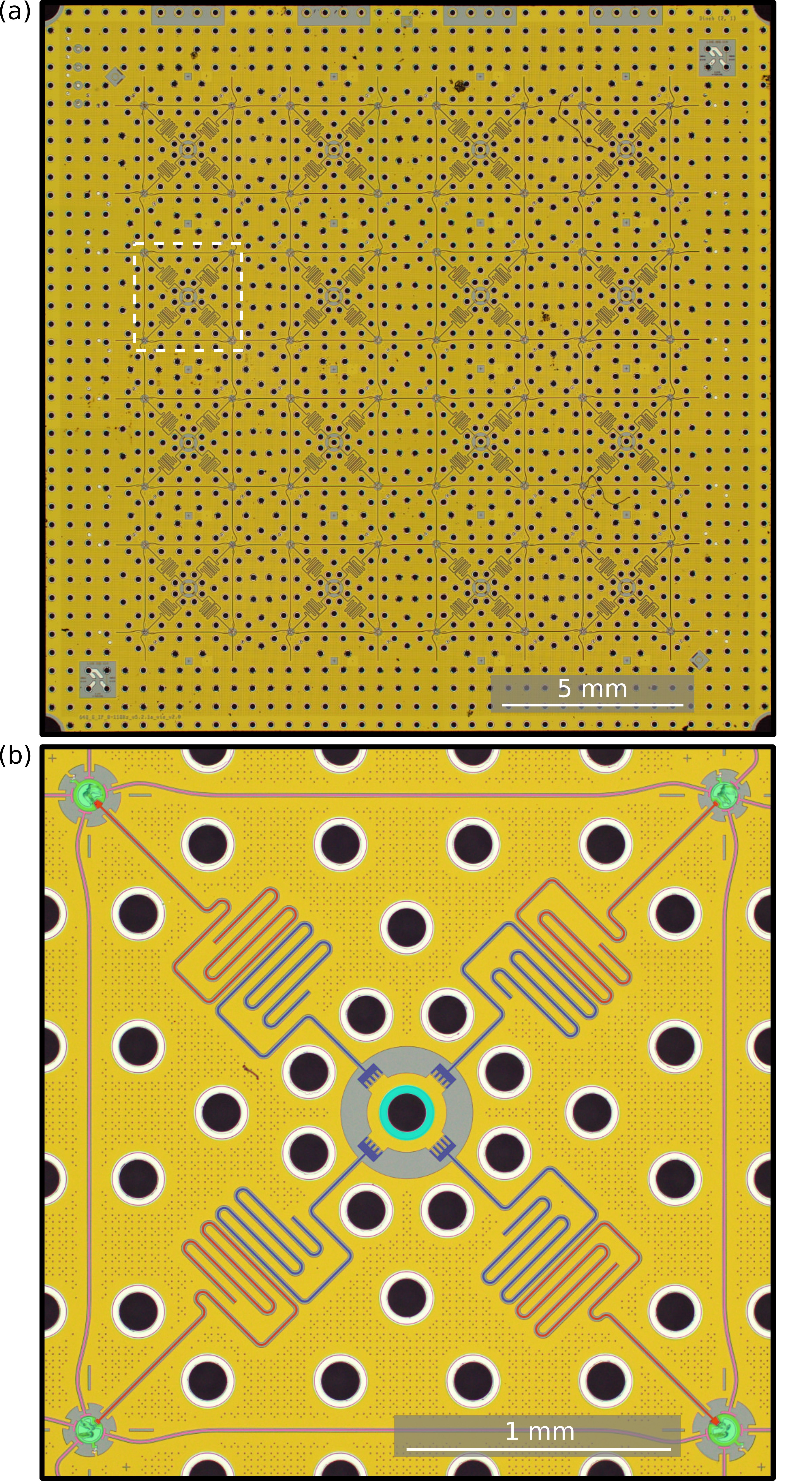}
\caption{\label{fig:Device_images} Images of the 64-qubit device. (a) Photograph of the 64-qubit chip. The unit cell containing the measured qubit is highlighted. (b) False-colored image of the four-qubit unit cell, highlighting the readout resonators~(red), filter resonators~(blue), qubits~(green), qubit--qubit couplers~(pink), and the TSV that connects to the readout feedline (teal). The dashed square highlights the qubit measured in this work.}
\end{figure}

\begin{figure}
\includegraphics[width=1\linewidth]{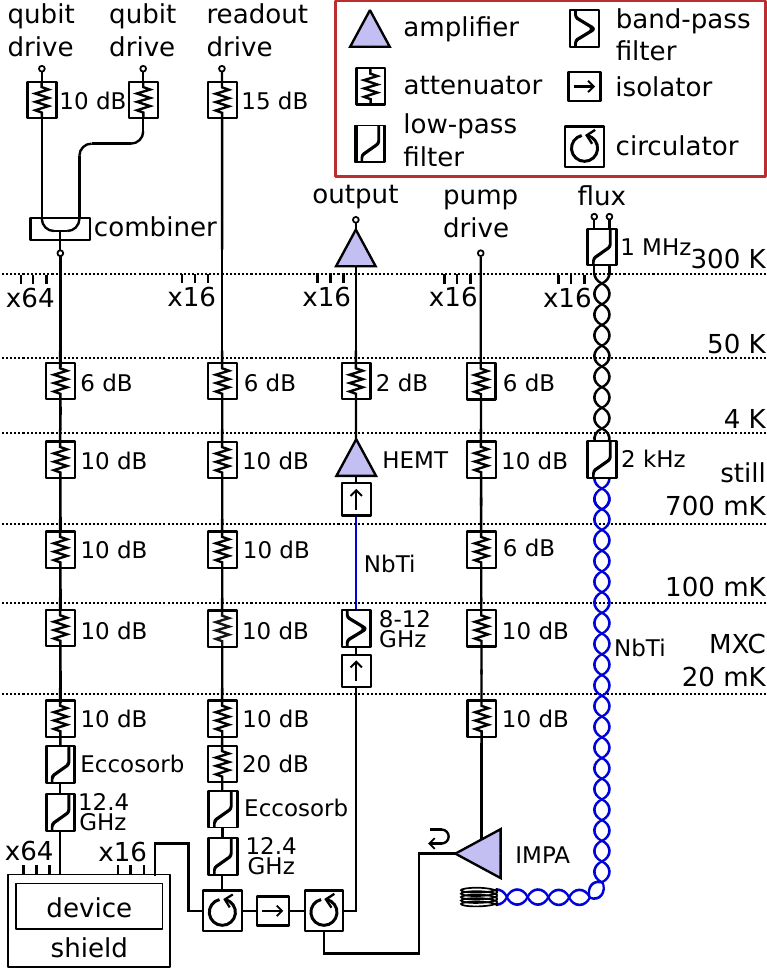}
\caption{\label{fig:Fridge_diagram} Experimental setup.}
\end{figure}

The qubit measured in this work is from the 64-qubit device shown in Fig.~\ref{fig:Device_images}(a). The qubit frequencies are tuned by Josephson-junction laser annealing~\cite{hertzberg2021laser}. Across the 52 operational qubits, the mean $T_1$ relaxation and Hahn-echo dephasing times are $\SI{19(9)}{\micro\second}$ and $\SI{23(8)}{\micro\second}$, respectively. Figure~\ref{fig:Device_images}(b) shows the four-qubit unit cell containing the measured qubit. The filter resonator couples to the readout resonator through a multiconductor-transmission-line section, which results in a notch filter that enhances Purcell filtering~\cite{spring_fast_2025}. The readout feedline contacts the backside of the chip, and the signal is routed to the filter resonators through a superconducting through-silicon via (TSV) metalized with Al~\cite{spring_fast_2025}. The Purcell-limited relaxation time imposed by the readout feedline predicted from a COMSOL simulation following Ref.~\citenum{sunada_fast_2022} is $\SI{3.5}{\second}$, as the qubit frequency is closely aligned to the notch frequency. The qubit parameters are summarized in Table~\ref{tab:qubit_extra_parameters_appendix}.
\\
\indent The device is cooled to approximately \SI{20}{\milli\kelvin} inside a
Bluefors XLD1000 dilution refrigerator. The experimental setup is shown in Fig.~\ref{fig:Fridge_diagram}. A QuEL-1 controller (QuEL Inc.) generates the qubit and readout drives and the impedance-matched parametric amplifier (IMPA) pump, and acquires the readout signal~\cite{machino2026qube}. The output of the qubit drive port is combined with that of an auxiliary port using a Wilkinson combiner. The auxiliary port is used to generate the $\uppi_{34}$ pulse for $|4\rangle$-state preparation.
\begin{table}[b]
\caption{\label{tab:qubit_extra_parameters_appendix}
Qubit parameters. The term $p_1$ is the excited state population of the qubit prior to state preparation. The Purcell-limited relaxation time is determined from a COMSOL simulation. All other quantities are experimentally determined.}
\begin{ruledtabular}
\begin{tabular}{lc}
 & $\mathrm{Q}_{16}$ \\
\colrule
$|0\rangle \to |1\rangle$ transition frequency $\omega_{\mathrm{q},01}/2\pi$ (\si{\mega\hertz}) & 8259.7 \\
$|1\rangle \to |2\rangle$ transition frequency $\omega_{\mathrm{q},12}/2\pi$ (\si{\mega\hertz}) & 7899.5 \\
$|2\rangle \to |3\rangle$ transition frequency $\omega_{\mathrm{q},23}/2\pi$ (\si{\mega\hertz}) & 7497.2 \\
$|3\rangle \to |4\rangle$ transition frequency $\omega_{\mathrm{q},34}/2\pi$ (\si{\mega\hertz}) & 7039.4 \\
Relaxation time $T_1$ (\si{\micro\second}) & 19 \\
Ramsey dephasing time $T_2^*$ (\si{\micro\second}) & 19 \\
Hahn-echo dephasing time $T_{2}^\mathrm{echo}$ (\si{\micro\second}) & 26 \\
Pure dephasing time $T_{\phi}^\mathrm{echo}$ (\si{\micro\second}) & 81 \\
Purcell-limited relaxation time $T_{1}^\mathrm{Purcell}$ (\si{\second}) & 3.5 \\
$|1\rangle$-state population $p_1$ (\%) & 0.35(3) \\
\end{tabular}
\end{ruledtabular}
\end{table}

\section{\label{appendix:input_output}Input--output model for readout dynamics}
\begin{figure}
\includegraphics[width=1\linewidth]{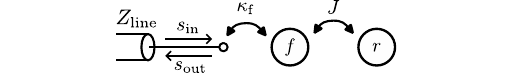}
\caption{\label{fig:input_output_model} Input--output network for the readout system.}
\end{figure}

This appendix describes the input--output model used to simulate the readout dynamics. The model is used to predict the steady-state and time-dependent responses to an external drive pulse, and to predict the mean integrated signals of heterodyne measurements.
\\
\indent Figure~\ref{fig:input_output_model} shows the input--output network for the readout system. A coherent traveling field $s_\mathrm{in}$ propagates along a waveguide with characteristic impedance $Z_\mathrm{line}$ and is reflected at the waveguide termination, producing the outgoing field $s_\mathrm{out}$. At the termination, the waveguide couples with coupling strength $\kappa_\mathrm{f}$ to the filter resonator, which in turn couples with coupling strength $J$ to the readout resonator. The coherent mode amplitudes in the filter and readout resonators are denoted by $f$ and $r$, respectively. The filter and readout resonator photon numbers, $n_\mathrm{f}$ and $n_\mathrm{r}$, respectively, are related to these mode amplitudes through the expressions $n_\mathrm{f} = |f|^2$ and $n_\mathrm{r} = |r|^2$. The ingoing and outgoing fields are related to the filter-mode amplitude by~\cite{gardiner1985input}
\begin{equation}
s_{\textrm{out}} = s_{\textrm{in}} - \sqrt{\kappa_\mathrm{f}}\, f \textrm{.} \label{eq:incoming outgoing field relation - appendix} 
\end{equation}
The equations of motion for the readout system can be represented by the matrix equation,
\begin{equation}
\label{eq:coupled_resonators_first_order_ode}
\frac{d}{dt}
\begin{pmatrix}
f \\
r
\end{pmatrix}
= 
-i \begin{pmatrix}
F & J \\
J & R^{n}\!
 \end{pmatrix}
\begin{pmatrix}
f \\
r
\end{pmatrix}
+
\begin{pmatrix}
\mathcal{E} \\
0
\end{pmatrix}
\textrm{,}
\end{equation}
where we define the drive amplitude $\mathcal{E} \equiv \sqrt{\kappa_\mathrm{f}} \, s_\mathrm{in}$, as  well as the terms,  
\begin{align}
F & \equiv \Delta_\mathrm{fd} - i\frac{\kappa_\mathrm{f}}{2} - i\frac{\gamma_\mathrm{f}}{2} \, \mathrm{,} \\
R^n & \equiv \Delta_\mathrm{rd}^n - i\frac{\gamma_\mathrm{r}}{2} \, \mathrm{.}
\end{align}
Here, $\Delta_\mathrm{fd}=\omega_\mathrm{f} - \omega_\mathrm{d}$ is the detuning of the filter resonator from the drive and $\Delta_\mathrm{rd}^n=\omega_\mathrm{r}^n - \omega_\mathrm{d}$ is the detuning of the readout resonator from the drive, with $\omega_\mathrm{r}^n$ being the readout-resonator frequency given that the qubit is in state $|n\rangle$. The quantities $\gamma_\mathrm{r}$ and $\gamma_\mathrm{f}$ are the internal decay rates of the readout and filter resonators, respectively.
\subsection{\label{appendix:B1subsec}Steady-state results}
We restate the definition for the SNR of the measurement given in Eq.~\eqref{eq:SNR_definition} in a squared form below, 
\begin{equation}
    \mathrm{SNR}^2 = 2\eta \int_0^{\tau} \left|s_\mathrm{out}^1(t) - s_\mathrm{out}^0(t) \right|^2 \mathrm{d}t \mathrm{.}
\end{equation}
In the steady state, using Eq.~\eqref{eq:incoming outgoing field relation - appendix}, this can be expressed in terms of the steady-state filter field $f^n$~($n=0,1$), as
\begin{equation}
    \mathrm{SNR}^2 = 2\eta\kappa_\mathrm{f} \tau \left|f^1 - f^0 \right|^2 \mathrm{.}
\end{equation}
The $\mathrm{SNR}^2$ per readout photon is given by 
\begin{equation}
    \mathrm{SNR}^2/\mathrm{avg}\left(n_\mathrm{r}\right) = 2\eta\kappa_\mathrm{f}\tau\frac{2\left|f^1 - f^0 \right|^2}{\left|r^0\right|^2+\left|r^1\right|^2}  \mathrm{,}
    \label{eq:SNR per photon - appendix}
\end{equation}
where $\mathrm{avg}\left(n_\mathrm{r}\right)\equiv \left(n_\mathrm{r}^0 + n_\mathrm{r}^1\right)/2$ and $r^n$ ($n=0, 1$) is the readout-resonator field. Solving Eq.~\eqref{eq:coupled_resonators_first_order_ode} in the steady state, under the conditions $\Delta_\mathrm{rd}=\Delta_\mathrm{rf}=0$, yields
\begin{equation}
    \mathrm{SNR}^2/\mathrm{avg}\left(n_\mathrm{r}\right) = 8 \eta \tau \kappa_\mathrm{f} \chi^2 \frac{J^2}{J^4 + \left(\kappa_\mathrm{f}\chi/2\right)^2} \, \mathrm{.}
    \label{eq:SNR_with_filter - appendix}
\end{equation}
The readout photon number satisfies $n_\mathrm{r}^0=n_\mathrm{r}^1$ for the special case $\Delta_\mathrm{rd}=\Delta_\mathrm{rf}=0$, so we can substitute $\mathrm{avg}\left(n_\mathrm{r}\right)\rightarrow n_\mathrm{r}$. Equation~\eqref{eq:SNR_with_filter - appendix} is maximized under the condition $\kappa_\mathrm{f}=2J^2/\chi$, in which case $\mathrm{SNR}^2_\mathrm{max}/n_\mathrm{r} = 8\eta|\chi|\tau$.
\\
\indent In the main text, we extract the readout parameters by fitting the measured qubit Stark shift $\chi_\mathrm{ac}$ characterized in the ``chi-kappa-power'' experiment, shown in Fig.~\ref{fig:parameter_characterization}(d), to a steady-state solution for the qubit Stark shift against readout drive frequency. Solving Eq.~\eqref{eq:coupled_resonators_first_order_ode} in the steady state yields the following qubit-state-dependent expression for the readout photon number $n_\mathrm{r}$ as a function of the drive frequency, 
\begin{equation}
    n_\mathrm{r}^n = \left|\frac{J \mathcal{E}}{J^2 - \left(\Delta_\mathrm{fd} - i\frac{\kappa_\mathrm{f}}{2} - i\frac{\gamma_\mathrm{f}}{2}\right)\left(\Delta_\mathrm{rd}^n - i\frac{\gamma_\mathrm{r}}{2} \right) }\right|^2 \, \mathrm{.} 
    \label{eq:readout_photon_number_vs_drive_frequency}
\end{equation}
The measured Stark shift $\chi_\mathrm{ac}$ is then fit using the relation $\chi_\mathrm{ac} = 2\chi n_\mathrm{r}^n$ for the $|0\rangle$ and $|1\rangle$ state. We set the internal decay rates $\gamma_\mathrm{r}$ and $\gamma_\mathrm{f}$ to zero when fitting since the large external decay rate $\kappa_\mathrm{f}$ dominates over the internal loss channels, resulting in Eq.~\eqref{eq:qubit_stark_shift_vs_drive_frequency}.
\subsection{\label{appendix:B2subsec}Normal modes}
In the absence of a drive, the two normal modes for the coupled readout- and filter-resonator system are found by diagonalizing Eq.~\eqref{eq:coupled_resonators_first_order_ode} in the steady state. The normal mode frequencies and linewidths can be expressed in closed form~\cite{swiadek_enhancing_2024}
\begin{align}
    \tilde{\omega}_\mathrm{r(f)} & = \frac{\omega_\mathrm{r}^n + \omega_\mathrm{f}}{2} +(-) \frac{1}{2}\mathrm{Re}\left[ \sqrt{\left(\Delta_\mathrm{rf}^n+i\kappa_\mathrm{f}/2\right)^2 + 4J^2}\right] \mathrm{,} \label{eq:omega_hybrid}\\
    \tilde{\kappa}_{\mathrm{r(f)}} & = \frac{\kappa_\mathrm{f}}{2} -(+) \mathrm{Im}\left[ \sqrt{\left(\Delta_\mathrm{rf}^n+i\kappa_\mathrm{f}/2\right)^2 + 4J^2}\right] \mathrm{,} \label{eq:kappa_hybrid}
\end{align}
where we neglect the internal decay rates $\gamma_\mathrm{r}$ and $\gamma_\mathrm{f}$. We define the normal mode with the larger decay rate as the filter-like mode at frequency $\tilde{\omega}_\mathrm{f}$, and the normal mode with the smaller decay rate as the readout-like mode at $\tilde{\omega}_\mathrm{r}$. The decay rates satisfy the relation ${\tilde{\kappa}_\mathrm{f} + \tilde{\kappa}_\mathrm{r}=\kappa_\mathrm{f}}$. These normal modes are used in the subsequent time-dependent results section.
\subsection{\label{appendix:B3subsec}Time-dependent results}
\begin{figure}
\includegraphics[width=1\linewidth]{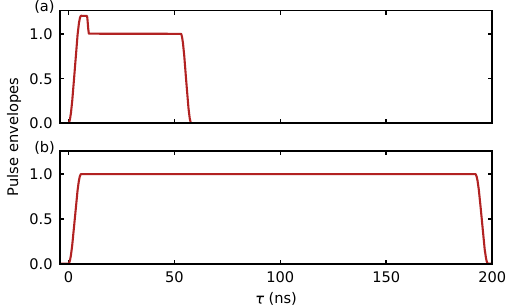}
\caption{\label{fig:measurement_pulse_envelopes} Measurement pulse envelopes. (a) Two-state measurement pulse envelope $u(\tau)$. (b) Leakage-sensitive measurement pulse envelope $u_\mathrm{leak}(\tau)$.}
\end{figure}

\begin{figure}
\includegraphics[width=1\linewidth]{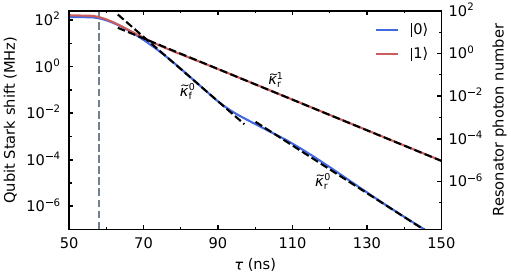}
\caption{\label{fig:ringdown_dynamics_appendix} Simulated readout photon number time dependence during ringdown. The two-state measurement pulse ends at $\SI{58}{\nano\second}$, indicated by the gray dashed line. The photon number after the drive switches off follows the sum of two exponentials expression given in Eq.~\eqref{eq:photon_number_solution_appendix}. When the qubit is in $|1\rangle$, the ringdown quickly settles on simple exponential decay at rate $\tilde{\kappa}_\mathrm{r}^1$, as indicated by the black-dashed line. When prepared in $|0\rangle$, the ringdown initially decays at rate $\tilde{\kappa}_\mathrm{f}^0$, before settling at the slower rate $\tilde{\kappa}_\mathrm{r}^0$.}
\end{figure}

We obtain the time-dependent resonator photon number by numerically integrating
Eq.~\eqref{eq:coupled_resonators_first_order_ode} under the drive
$\mathcal{E}(\tau)=\mathcal{E}_0\,u(\tau)$, where $u(\tau)$ is the two-state
measurement-pulse envelope of Fig.~\ref{fig:measurement_pulse_envelopes}(a).
Using the relation $\chi_\mathrm{ac}(\tau)=2\chi\,n_\mathrm{r}^n(\tau)$ for the $|0\rangle$ and $|1\rangle$ state, we fit the computed photon number to the measured qubit Stark-shift data in
Fig.~\ref{fig:resonator_time_response}(b), treating the drive amplitude
$\mathcal{E}_0$ as a free parameter and fixing the remaining readout
parameters to the predetermined values from the ``chi-kappa-power'' experiment. The fit yields $\mathcal{E}_0/2\pi=\SI{82.7}{\mega\hertz}$. From the resulting readout- and filter-mode amplitudes, we evaluate the state-dependent outgoing field $s^n_\mathrm{out}(\tau)$ using Eq.~\eqref{eq:incoming outgoing field relation - appendix} and plot the predicted output-field separation $\Delta s_\mathrm{out}(\tau)=\left|s_\mathrm{out}^0(\tau)-s_\mathrm{out}^1(\tau)\right|$ in Fig.~\ref{fig:two_state_measurement}(b).
\\
\indent After the drive switches off, the resonator field $r$ evolves as a superposition of the two hybridized modes, which decay at rates $\tilde{\kappa}_\mathrm{r}$ and $\tilde{\kappa}_\mathrm{f}$. Solving Eq.~\eqref{eq:coupled_resonators_first_order_ode} with no drive, initial field amplitudes $r_0$ and $f_0$, and vanishing internal loss ($\gamma_\mathrm{r}=\gamma_\mathrm{f}=0$), yields the following solution for the readout-photon-number decay,
\begin{multline}
    n_\mathrm{r}(\tau) = |\mathcal{A}|^2e^{-\tilde{\kappa}_\mathrm{f} \tau} + |\mathcal{B}|^2e^{-\tilde{\kappa}_\mathrm{r} \tau} \\ + 2|\mathcal{A}\mathcal{B}|e^{-\left(\kappa_\mathrm{f}/2\right) \tau }\cos[\left(\tilde{\omega}_\mathrm{f} - \tilde{\omega}_\mathrm{r} \right)\tau - \phi] \mathrm{,}
    \label{eq:photon_number_solution_appendix}
\end{multline}
with the relations
\begin{align} 
    \mathcal{A} & = \frac{Jf_0 + \left( \omega_\mathrm{r}^n - \tilde{\omega}_\mathrm{r} + i\tilde{\kappa}_\mathrm{r}/2 \right)r_0}{\tilde{\omega}_\mathrm{f} - \tilde{\omega}_\mathrm{r} - i\left(\tilde{\kappa}_\mathrm{f} - \tilde{\kappa}_\mathrm{r}\right)/2} \mathrm{,} \\
    \mathcal{B} & =  -\frac{Jf_0 + \left( \omega_\mathrm{r}^n - \tilde{\omega}_\mathrm{f} + i\tilde{\kappa}_\mathrm{f}/2  \right)r_0}{\tilde{\omega}_\mathrm{f} - \tilde{\omega}_\mathrm{r} - i\left(\tilde{\kappa}_\mathrm{f} - \tilde{\kappa}_\mathrm{r}\right)/2} \mathrm{,} \\
    \phi & = \mathrm{arg}\left(\mathcal{A}\mathcal{B}^*\right) \mathrm{.}
\end{align}
The readout photon number is the sum of two exponentials, decaying at the hybridized-mode decay rates $\tilde{\kappa}_\mathrm{r}$ and $\tilde{\kappa}_\mathrm{f}$, with an additional interference term that decays at the rate $\kappa_\mathrm{f}/2$. At long times the decay is set by the slower, readout-like mode as
\begin{equation}
    n_\mathrm{r}(\tau) \approx  |\mathcal{B}|^2e^{-\tilde{\kappa}_\mathrm{r} \tau} \mathrm{.}
\end{equation}
\indent Figure~\ref{fig:ringdown_dynamics_appendix} shows the photon-number dynamics predicted by the input--output model following the two-state measurement pulse. The drive ends at $\tau=\SI{58}{\nano\second}$, beyond which the photon number $n_\mathrm{r}(\tau)$ follows Eq.~\eqref{eq:photon_number_solution_appendix}. For the qubit in $|1\rangle$, the hybridized mode decay rates are widely separated, with $\tilde{\kappa}^1_\mathrm{r}/2\pi=\SI{24.1}{\mega\hertz}$ and $\tilde{\kappa}^1_\mathrm{f}/2\pi=\SI{64.2}{\mega\hertz}$. As a result, the ring down settles rapidly onto a single exponential with decay rate $\tilde{\kappa}^1_\mathrm{r}/2\pi$. For the qubit in $|0\rangle$, the hybridization is stronger and the decay rates lie closer together, with $\tilde{\kappa}^0_\mathrm{r}/2\pi=\SI{37.5}{\mega\hertz}$ and $\tilde{\kappa}^0_\mathrm{f}/2\pi=\SI{50.8}{\mega\hertz}$. In this case, it takes longer for the photon number to ring down at the rate $\tilde{\kappa}_\mathrm{r}^0$.
At the earlier times, the decay rate follows the faster decaying hybridized mode, $\tilde{\kappa}_\mathrm{f}^0/2\pi = \SI{50.8}{\mega\hertz}$, leading to a very rapid initial decay of the readout photon number. The readout-like-mode decay rate therefore sets a conservative lower bound on the decay of readout photons.
\subsection{\label{appendix:B4subsec}Integrated readout signals}
The time-dependent outgoing fields $s_\mathrm{out}$ can be used to predict the relative position of the weighted integrated readout signals in the IQ plane. We define the integrated outgoing field signal, $S^n_\mathrm{out}$, given the qubit in state $|n\rangle$, as
\begin{align}
S^n_\mathrm{out} = \int_0 ^{t_\mathrm{int}} w_\mathrm{out}\!\left(\tau\right) s^n_\mathrm{out}\!\left(\tau\right) \mathrm{d}\tau \textrm{,} 
\label{eq:outgoing integrated field}
\\
w_\mathrm{out}\!\left(\tau\right) \equiv \left[s_\mathrm{out}^{0}\!\left(\tau\right) - s_\mathrm{out}^{1}\!\left(\tau\right)\right]^{*} \, \textrm{.}
\label{eq:outgoing field weighting function}
\end{align}
Matched-filter weighting is used in order to simulate the measured integrated signals. In Fig.~\ref{fig:two_state_measurement}(c), the predicted integrated signals in response to the two-state measurement pulse, using a integration window $t_\mathrm{int}$ of \SI{72}{\nano\second}, are overlaid on the measured integrated signals. The measured and predicted signals are normalized by the $|0\rangle$-state signal magnitudes $|\overline{S}^0|$ and $|S_\mathrm{out}^0|$, respectively, which enables direct comparison without any fit parameters. The same approach is used to predict the integrated signals for the leakage-sensitive measurement, as described in more detail in Appendix~\ref{appendix:leakage_sensitive_measurement}.
\section{\label{appendix:quantum_model} Skewed distribution in chi-kappa-power experiment}
\begin{figure}
\includegraphics[width=1\linewidth]{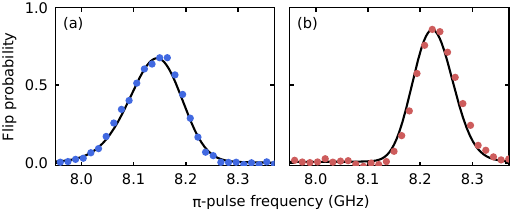}
\caption{\label{fig:Kappa_Chi_Power_sim} Investigation of skewed distribution in chi-kappa-power experiment. (a),(b)~Measured qubit-state flip probability versus $\uppi$-pulse frequency at readout-drive carrier frequency $\SI{10248}{\mega\hertz}$, for the qubit prepared in $|0\rangle$ and $|1\rangle$, respectively. The black curves shows the prediction of the time-dependent simulation.}
\end{figure}
\indent We observe that the qubit-state-flip probability versus $\uppi$-pulse carrier frequency in the chi-kappa-power experiment exhibits a mildly skewed distribution. Figure~\ref{fig:Kappa_Chi_Power_sim} shows the measured flip probability against $\uppi$-pulse frequency for the qubit prepared in $|0\rangle$ and $|1\rangle$, at readout-drive carrier frequency $\SI{10248}{\mega\hertz}$. Both distributions exhibit a skew, with the skew having opposite sign for the qubit prepared in $|0\rangle$ and $|1\rangle$. To investigate this skew, we simulate the experiment by
numerically solving the Lindblad master equation for the chi-kappa-power pulse sequence. The qubit is treated as a four level system, the readout resonator is truncated to 20 levels, and the filter resonator is adiabatically eliminated~\cite{sete_quantum_2015}. The Hamiltonian parameters are determined by fitting to the qubit and readout parameters in Table~\ref{tab:qubit_readout_parameters}. The black curves show the prediction of the simulation. The simulation reproduces the scale and direction of the skew, indicating that it is a feature of the driven qubit--readout-resonator dynamics. Investigating the mechanism further is beyond the scope of this work.
\section{\label{appendix:total_measurement_duration}Total Measurement Duration}
In this Appendix, we derive the error model used to infer the added error to the $\uppi$ pulse from residual photons in the readout resonator. We also describe the pulse sequence used to verify the qubit-readout pulse timing calibration.
\subsection{\label{appendix:D1subsec} Error-model derivation}
\begin{figure}
\includegraphics[width=1\linewidth]{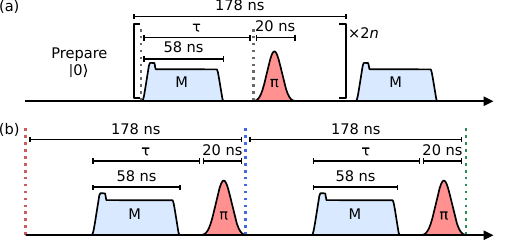}
\caption{\label{fig:total_readout_duration_error_model_appendix} Total-readout-duration characterization. (a)~Pulse sequence for the experiment, reproduced from Fig.~\ref{fig:total_measurement_duration}. (b)~Pulse sequence for two repeated measurement--$\uppi$-pulse blocks. We construct the error model by tracking the qubit state at the red, blue, and green dotted lines.}
\end{figure}

The pulse sequence used to determine the total measurement duration in the main text is reproduced in Fig.~\ref{fig:total_readout_duration_error_model_appendix}(a). After the qubit is prepared in $|0\rangle$, the measurement--$\uppi$-pulse block is repeated $2n$ times, followed by a final measurement. In Fig.~\ref{fig:total_readout_duration_error_model_appendix}(b), the pulse sequence for two repeated blocks is shown explicitly. We use this sequence to construct the error model. Given that the qubit is in $|0\rangle$ at the red dotted line, the probability that it remains in $|0\rangle$ directly after the first $\uppi$ pulse~(blue dotted line) is given by
\begin{multline}
  \mathcal{P}(0|0) = \epsilon_{\uparrow} \left[1-\epsilon_\uppi^1(\tau) - \epsilon_{\uppi,0}\right] \\
  + \left[1-\epsilon_{\uparrow} \right]\left[\epsilon_\uppi^0(\tau) + \epsilon_{\uppi,0}\right] \mathrm{.}
\end{multline}
Here, $\epsilon_\uppi^0(\tau)[\epsilon_\uppi^1(\tau)]$ is the time-dependent $\uppi$-pulse error due to residual photons in the resonator, given that the qubit is in $|0\rangle[|1\rangle]$
during the preceding measurement. The term $\epsilon_{\uppi,0}$ is the time-independent $\uppi$-pulse error unrelated to residual resonator photons. The term $\epsilon_\uparrow$ is the $|0\rangle$$\rightarrow$$|1\rangle$ relaxation error during the ${178}$-\si{\nano\second} interval between the red and blue dotted lines. Note that we drop the time dependence of $\mathcal{P}$ for readability. 
\\
\indent Given that the qubit is instead in $|1\rangle$ at the red dotted line, the probability that it remains in $|1\rangle$ at the blue dotted line is given by
\begin{multline}
   \mathcal{P}(1|1) = \epsilon_\downarrow \left[1-\epsilon_\uppi^0(\tau) - \epsilon_{\uppi,0} \right] \\
   + \left[1-\epsilon_\downarrow\right]\left[\epsilon_\uppi^1(\tau) + \epsilon_{\uppi,0} \right]\mathrm{.}
\end{multline}
Here, $\epsilon_\downarrow$ is the $|1\rangle$$\rightarrow$$|0\rangle$ relaxation error during the $178$\nobreakdash-\si{\nano\second} interval between the red and blue dotted lines.
\\
\indent We now consider the qubit state at the end of the second measurement--$\uppi$-pulse block. Given that the qubit is in $|0\rangle$ at the red dotted line, the probability that it has switched to $|1\rangle$ at the end of the second $\uppi$-pulse (green dotted line) is given by
\begin{equation}
    \epsilon_{01}(\tau) =  \mathcal{P}(1|1) \mathcal{P}(1|0) + \mathcal{P}(1|0) \mathcal{P}(0|0)  \mathrm{.}
\end{equation}
Using the identity $\mathcal{P}(1|0) = 1 - \mathcal{P}(0|0)$ and expanding yields
\begin{align}
    \epsilon_{0 1}(\tau) & = 2\epsilon_\uppi(\tau) + 2\epsilon_{\uppi,0} + \epsilon_{\uparrow \downarrow} + \mathcal{O}\left(\epsilon^2\right) \mathrm{.}
    \label{eq:error_0_to_1}
\end{align}
Here, we have defined the state-averaged added $\uppi$-pulse error $\epsilon_\uppi(\tau) \equiv \left[\epsilon_\uppi^0(\tau) + \epsilon_\uppi^1(\tau)\right]/2$, and the total relaxation error $\epsilon_{\uparrow \downarrow} \equiv \epsilon_{\uparrow } + \epsilon_\downarrow$. 
\\
\indent If instead the qubit is in $|1\rangle$ at the red dotted line, the probability that it has switched to $|0\rangle$ at the green dotted line is given by
\begin{equation}
    \epsilon_{10}(\tau) =  \mathcal{P}(0|1) \mathcal{P}(1|1) + \mathcal{P}(0|0) \mathcal{P}(0|1)  \mathrm{.}
    \label{eq:error_1_to_0}
\end{equation}
Using the identity $\mathcal{P}(0|1) = 1 - \mathcal{P}(1|1)$ and expanding yields
\begin{align}
    \epsilon_{10}(\tau) & = 2\epsilon_\uppi(\tau) + 2\epsilon_{\uppi,0} + \epsilon_{\uparrow \downarrow} + \mathcal{O}\left(\epsilon^2\right) \mathrm{.}
\end{align}
\indent The pulse sequence in Fig.~\ref{fig:total_readout_duration_error_model_appendix}(b) is repeated $n$ times with the qubit initially prepared in $|0\rangle$. The probability $P_1$ that qubit is in $|1\rangle$ immediately prior to the final measurement is then given by the two-state Markov-chain solution:
\begin{align}
P_1 & = \alpha \left (1-\gamma^n \right) \mathrm{,} \\
\alpha & \equiv \frac{\epsilon_{01}}{\epsilon_{10} + \epsilon_{01}} \mathrm{,} \\
\gamma & \equiv 1 - \epsilon_{01} - \epsilon_{10} \mathrm{.}
\end{align}
The flip errors $\epsilon_{1 0}(\tau)$ and $\epsilon_{0 1}(\tau)$ are equal to leading order, and we define the leading-order term $\epsilon(\tau)\equiv 2\epsilon_\uppi(\tau) + 2\epsilon_{\uppi,0} + \epsilon_{\uparrow \downarrow}$. The difference between $\epsilon_{1 0}(\tau)$ and $\epsilon_{0 1}(\tau)$ can be expressed in terms of $\epsilon(\tau)$ as
\begin{multline}
    \epsilon_{0  1}(\tau) - \epsilon_{1 0}(\tau) = \left[\epsilon_\downarrow - \epsilon_\uparrow +\epsilon_{\uppi}^1(\tau) - \epsilon_{\uppi}^0(\tau) \right]\epsilon(\tau) \\ + \mathcal{O}\left(\epsilon^3\right) \mathrm{.} 
\end{multline}
This difference is small for our data, where the fitted values of $\epsilon(\tau)$ lie between $1\%$ and $7.5\%$. Therefore, we make the approximation $\epsilon_{10}(\tau)= \epsilon_{0 1}(\tau)= \epsilon(\tau)$, which yields the single-fit-parameter solution 
\begin{equation}
    P_1 = \{1 - \left[1-2\epsilon(\tau)\right]^n \}/2 \mathrm{.} \label{eq:added_pi_error_model - appendix} 
\end{equation}
\indent Allowing for an $n$-independent state preparation and measurement error then results in Eq.~\eqref{eq:added_pi_error_model}. 
\subsection{\label{appendix:D2subsec} Verifying the calibration of timing $\tau$}
\begin{figure}
\includegraphics[width=1\linewidth]{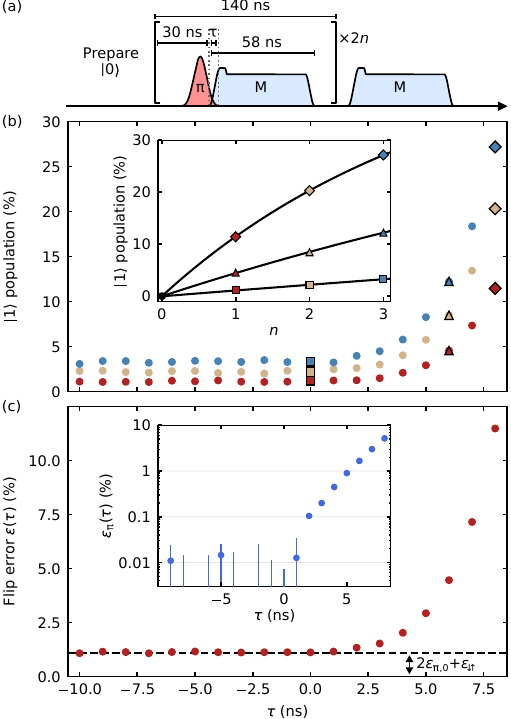}
\caption{\label{fig:readout_start_time_sequence_appendix} Verifying $\tau$ calibration. (a) Pulse sequence. The time $\tau$ is defined as the interval between the measurement start and the $\uppi$-pulse end. The duration of the $\uppi$ pulse is \SI{20}{\nano\second}. The $\uppi$-pulse--measurement block is repeated $2n$ times. (b)~Excited-state population determined by the final measurement pulse, for $n=1$ (red), $n=2$ (tan), and $n=3$ (blue). The inset shows the excited-state population for three fixed values of $\tau$. The corresponding data points are indicated with black open markers in the main panel. In the inset, the black circle at $n=0$ is the independently-measured SPAM error $\epsilon_\textrm{SPAM}$, and the black lines show the fit to Eq.~\eqref{eq:added_pi_error_model}. (c) Flip error $\epsilon(\tau)$ extracted from (b). The $\tau$-independent error is shown by the black dashed line. The inset shows the inferred $\uppi$-pulse error due to overlap with the measurement pulse.}
\end{figure}

The total measurement duration $t_\mathrm{meas}$ is defined as the time from the start of the measurement pulse until the measurement-induced error on a subsequent $\uppi$-pulse operation falls below a threshold value. In the main text, for the threshold value $10^{-4}$ we infer a $t_\mathrm{meas}$ value of $\SI{97(1)}{\nano\second}$, using the pulse sequence shown in Fig.~\ref{fig:total_readout_duration_error_model_appendix} to benchmark the error of a $\uppi$ pulse commencing at time $\tau$ after the start of the measurement pulse. Here, we use a complementary pulse sequence to verify the calibration of $\tau$, by checking that a $\uppi$ pulse performed before the measurement, terminating at time $\tau=0$, is not degraded by the measurement. The pulse sequence is shown in Fig.~\ref{fig:readout_start_time_sequence_appendix}(a). In this sequence, the $\uppi$ pulse precedes the measurement pulse, and the time $\tau$ is swept such that the end of the $\uppi$ pulse overlaps with the beginning of the measurement pulse. 
Figure~\ref{fig:readout_start_time_sequence_appendix}(b) shows the measured final $|1\rangle$-state population for different values of the repetition number $n$. The same error model as for the ring-down applies, and we fit the final excited-state population to Eq.~\eqref{eq:added_pi_error_model}, as shown in the inset for particular values of $\tau$. Figure~\ref{fig:readout_start_time_sequence_appendix}(c) shows the fitted flip errors $\epsilon(\tau)$. For $\tau>0$, the error rises rapidly as the $\uppi$ pulse overlaps with the start of the measurement. No added error is detected for $\tau\leq\SI{0}{\nano\second}$, confirming that a $\uppi$ pulse ending at time $\tau=\SI{0}{\nano\second}$ is not degraded by a measurement pulse commencing at $\tau=\SI{0}{\nano\second}$. The background $B=2\epsilon_{\uppi,0} + \epsilon_{\uparrow\downarrow}$ is dominated by relaxation error, and the fitted value $B=0.0109$ corresponds to a relaxation time of $T_{1,\mathrm{eff}}=\SI{140}{\nano\second}/0.0109=\SI{12.8}{\micro\second}$, consistent with the value of $T_{1,\mathrm{eff}}=\SI{178}{\nano\second}/0.0141=\SI{12.6}{\micro\second}$ determined from the corresponding ringdown pulse sequence in Fig.~\ref{fig:total_measurement_duration}.
\section{\label{appendix:leakage_sensitive_measurement} Leakage-sensitive measurement}
\begin{figure*}
\includegraphics[width=1\linewidth]{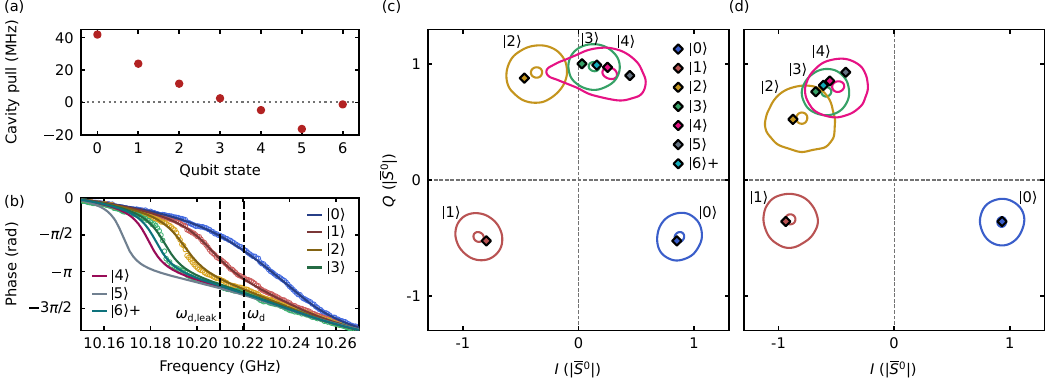}
\caption{\label{fig:leakage_sensitive_measurement_appendix}Dispersive shifts and readout response for higher transmon states. (a) Simulated cavity pull $\chi_n$ for the transmon in states $|0\rangle$--$|6\rangle$. (b) Phase response of the reflected readout signal. Hollow circles show the measured response for states $|0\rangle$--$|3\rangle$. Solid lines show the response predicted by the input--output model for states $|0\rangle$--$|6\rangle+$, using the simulated cavity pulls from~(a) and no adjustable parameters. Black dashed lines indicate the two drive frequencies, $\omega_\mathrm{d}$ for the two-state measurement and $\omega_{\mathrm{d},\mathrm{leak}}$ for the leakage-sensitive measurement. (c) Probability density contours at 5\% and 68\% of the measured integrated IQ signals for states $|0\rangle$--$|4\rangle$, for a readout drive at $\omega_{\mathrm{d},\mathrm{leak}}$. Diamond markers show the mean signals predicted by the input--output model for states $|0\rangle$--$|6\rangle+$, without adjustable parameters. (d) Same as (c), for a readout drive at $\omega_\mathrm{d}$.
}
\end{figure*}
\indent In this appendix, we describe how the input--output model can predict the relative position of the integrated IQ signals for leakage states that could not be prepared experimentally. We then describe the signature-based classification of the leakage-sensitive measurement data, and the thresholding technique we apply to reduce false-positive leakage detection.
\subsection{\label{appendix:E1subsec} Prediction of integrated signals with input--output model}
\indent For the leakage-sensitive measurement, it is desirable that all leakage-state signals are well separated from the $|0\rangle$- and $|1\rangle$-state signals. However, we could only prepare the qubit in states $|0\rangle$ to $|4\rangle$ due to the low coherence and charge dispersion of higher transmon levels. Nevertheless, we can use the input--output model to predict the position of the integrated signals in the IQ plane for higher-level qubit states $|n\rangle$ so long as we know the cavity pull $\chi_n \equiv \omega_\mathrm{r}^n - \omega_\mathrm{r}'$, where $\omega_\mathrm{r}'$ is the bare resonator frequency. We solve for the cavity pull using the Hamiltonian
\begin{equation}
\hat{H} = 4E_C\bigl(\hat{n} - n_g\bigr)^2 - E_J\cos\hat{\varphi} + \hbar \omega_\mathrm{r}'\, \hat{a}^\dagger \hat{a} + \hbar g\,\hat{n}\,\bigl(\hat{a} + \hat{a}^\dagger\bigr) \mathrm{,}
\end{equation}
where $E_C$ and $E_J$ are the qubit charging energy and Josephson energy, respectively, $n_\mathrm{g}$ is the gate charge, and $g$ is the qubit--resonator transverse coupling strength. The terms $\hat{n}$ and $\hat{\varphi}$ are the Cooper-pair number operator and phase operator of the transmon, and $\hat{a}^\dagger$ and $\hat{a}$ are the creation and
annihilation operators for the readout-resonator mode. From the measured qubit frequency, anharmonicity, resonator frequency and dispersive shift, we fit the Hamiltonian parameters $E_J$, $E_C$, $\omega_\mathrm{r}'$, and $g$. The fitted values are given in Table~\ref{tab:qubit_readout_Hamiltonian_parameters - appendix}. The predicted cavity pulls are shown in Fig.~\ref{fig:leakage_sensitive_measurement_appendix}(a) for qubit states up to $n=6$. The cavity pull for $n=6$ is small, at $\chi_6/2\pi=-\SI{1.2}{\mega\hertz}$, and for higher leakage states $n \geq 7$ the cavity pull tends to zero as the effect of the Josephson energy $E_J$ effectively vanishes for the highly excited transmon states~\cite{lescanne2019escape}. Thus, the readout resonator frequency for states $|6\rangle$ and above, denoted $|6\rangle+$, is close to the bare resonator frequency $\omega_\mathrm{r}'$.
\\
\indent In Fig.~\ref{fig:leakage_sensitive_measurement_appendix}(b), we plot the measured steady-state phase response $\mathrm{arg}(\Gamma_\mathrm{meas}$) of the reflected readout signal, for the qubit prepared in $|0\rangle$ to $|3\rangle$. The predicted phase response $\mathrm{arg}\left(\Gamma_\mathrm{sim}\right)$ is overlaid for the qubit prepared in $|0\rangle$ to $|6\rangle+$. The reflection coefficient $\Gamma_\mathrm{sim}=s_\mathrm{out}/s_\mathrm{in}$ is solved for using the input--output model described in Appendix~\ref{appendix:input_output}. For states $|0\rangle$--$|3\rangle$, the input--output model shows reasonably good correspondence with the data. The two-state measurement frequency $\omega_\mathrm{d}$ and the leakage-sensitive measurement frequency $\omega_{\mathrm{d},\mathrm{leak}}$ are indicated by the black dashed lines. At the leakage-sensitive measurement frequency, the $|0\rangle$-, $|1\rangle$- and $|2\rangle$-state phase responses are well separated, while the phase responses for the leakage states $|3\rangle$ and above are almost indistinguishable. Physically, this corresponds to the limit where the shifted resonator frequency is far-detuned from both the filter resonator and the drive frequency, and so further detuning has an increasingly weak effect on the output field. This implies that the integrated signals for the leakage states $|3\rangle$ and above will likewise be in close proximity in the IQ plane.
\begin{table}[b]
\caption{\label{tab:qubit_readout_Hamiltonian_parameters - appendix}
Qubit and readout-resonator Hamiltonian parameters.}
\begin{ruledtabular}
\begin{tabular}{ccccc}
$E_J/h$ & $E_C/h$ & $E_J/E_C$ & $\omega_r'/2\pi$ & $g/2\pi$ \\
\textrm{(MHz)}&
\textrm{(MHz)}&
\textrm{-} &
\textrm{(MHz)}&
\textrm{(MHz)}\\
\colrule
27487 & 342 & 80.5 & 10191 & 243 \\
\end{tabular}
\end{ruledtabular}
\end{table}
\\
\indent Figures~\ref{fig:leakage_sensitive_measurement_appendix}(c) and (d) show the measured integrated signals with the qubit prepared in states $|0\rangle$ to $|4\rangle$, for readout at $\omega_{\mathrm{d},\mathrm{leak}}/2\pi = \SI{10210}{\mega\hertz}$ and $\omega_\mathrm{d}/2\pi = \SI{10220.5}{\mega\hertz}$, respectively. Overlaid are the predicted signals for states $|0\rangle$ to $|6\rangle+$, obtained from the integrated output fields. We solve for the outgoing field $s_\mathrm{out}(\tau)$ in response to a measurement pulse with the time-dependent drive $\mathcal{E} = \mathcal{E}_0 u_\mathrm{leak}(\tau)$, where $u_\mathrm{leak}(\tau)$ is the flat-top envelope shown in Fig.~\ref{fig:measurement_pulse_envelopes}(b), applied at frequencies $\omega_{\mathrm{d},\mathrm{leak}}$ and $\omega_\mathrm{d}$, respectively. The field is integrated over a $t_\mathrm{int}=\SI{192}{\nano\second}$ window, matching the experiment, using the $|0\rangle$--$|1\rangle$ matched-filter weighting function in Eq.~\eqref{eq:outgoing field weighting function}. At both drive frequencies, the predicted signals for states $|0\rangle$ to $|4\rangle$ agree well with the measured signals without adjustable parameters. The predicted signals for the leakage states $|5\rangle$ and $|6\rangle+$ are in close proximity to the $|4\rangle$-state signal, so although these states could not be prepared directly, we expect them to be correctly identified as leakage by the leakage-sensitive measurement.
\subsection{\label{appendix:E2subsec} Signature-based classification and leakage thresholding}
\indent Given the acquired time-dependent signal $s^n(\tau)$ for the qubit in state $|n\rangle$, we form the path $X^n(t)$ defined in Eq.~\eqref{eq:integrated_path}, restated here for convenience,
\begin{equation}
        X^n(t) = \int_0^{t} w\!\left(\tau \right) s^n(\tau) \,\mathrm{d}\tau \textrm{.}
    \label{eq:integrated_path_appendix} 
\end{equation}
From $X^n(t)$ we construct the time-augmented log-signature~\cite{reizenstein2020algorithm, chevyrev2025primer, cao2024superconducting}, computed with the \texttt{iisignature} library~\cite{reizenstein2020algorithm}. The depth-1 log-signature reduces to the conventional matched-filter integrated signal $S$, corresponding to the two-element feature set $\{S_I, S_Q\}$. Note that for the depth-1 log-signature we drop the time-augmentation element $t$ because it contains no discriminating information. The higher-depth signatures form progressively larger supersets, with the feature set size growing rapidly with the signature depth $N$. For each depth, we split the data into a training set of $\num{1e4}$ shots per qubit state and a test set of $\num{5e4}$ shots per qubit state, and train a multinomial logistic regression (LR) classifier, yielding the assignment fidelity $\mathcal{F}_\mathrm{leak}$ as a function of depth. We restate the definition $\mathcal{F}_\mathrm{leak} \equiv \left[\mathcal{P}\left(0|0\right) +\mathcal{P}\left(1|1\right) + \mathcal{P}_4\!\left(\mathcal{L}|\mathcal{L}\right) \right]/3 $, with $\mathcal{P}_4\!\left(\mathcal{L}|\mathcal{L}\right) = \left[\mathcal{P}\left(\mathcal{L}|2\right) + \mathcal{P}\left(\mathcal{L}|3\right) + \mathcal{P}\left(\mathcal{L}|4\right) \right]/3 $.
\begin{figure}
\includegraphics[width=0.99\linewidth]{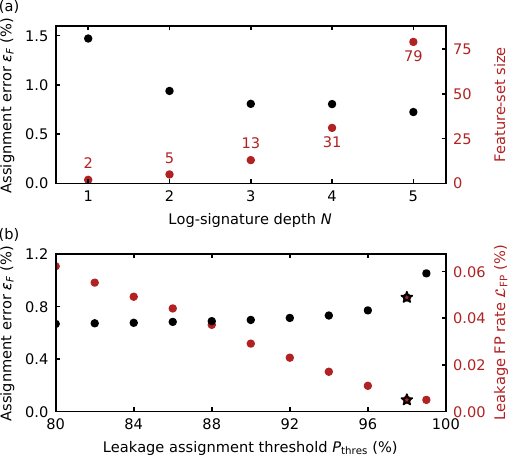}
\caption{\label{fig:leakage_sensitive_classifier_appendix} Classifier optimization for the leakage-sensitive readout. (a)~Assignment error $\epsilon_{\mathcal{F},\textrm{leak}}$ (left axis) and feature-set size (right axis) versus log-signature depth, without leakage thresholding. (b)~Assignment error (left axis) and leakage-false-positive rate $\mathcal{L}_\mathrm{FP}$ (right axis) versus the leakage threshold $P_\mathrm{thres}$, using the depth-5 log-signature. In~(b), star markers indicate the threshold used in the main text.}
\end{figure}
\\
\indent Figure~\ref{fig:leakage_sensitive_classifier_appendix}(a) shows the assignment error $\varepsilon_{\mathcal{F},\mathrm{leak}}\equiv 1 - \mathcal{F}_\mathrm{leak}$ and the feature-set size versus the log-signature depth. The assignment error decreases with increasing depth, from $\epsilon_{\mathcal{F},\textrm{leak}}=1.47\%$ at $N=1$ to $\epsilon_{\mathcal{F},\textrm{leak}}=0.72\%$ at $N=5$, representing a $50\%$ error reduction. We use the depth-5 log signature in the main text as it achieves the best assignment fidelity. We do not explore higher depths due to the rapid increase in the feature-set size.
\\
\indent In the MIST benchmarking experiment, we need to detect small leakage populations, which are obscured if the leakage false-positive (FP) detection rate is too high. To reduce leakage FPs, we  use a simple thresholding technique. For each shot, the multinomial LR model returns the predicted probability for each class, and by default assigns to the class with the highest predicted probability. We adjust this so that the model will only assign the $\mathcal{L}$ class when the predicted probability is above a threshold value $P_{\textrm{thres}}$. Otherwise, the model will assign the next most probable class. The leakage FP rate is defined as $\mathcal{L}_\textrm{FP} \equiv \left[\mathcal{P}(\mathcal{L}|0) + \mathcal{P}(\mathcal{L}|1) \right]\!/2$. Figure~\ref{fig:leakage_sensitive_classifier_appendix}(b) shows that as the threshold $P_{\textrm{thres}}$ for the $\mathcal{L}$ class increases, $\mathcal{L}_\textrm{FP}$ decreases approximately linearly, at the cost of a modest increase in the assignment error. We use the threshold $P_\textrm{thres}= 98 \%$, which results in a low leakage FP rate $\mathcal{L}_\textrm{FP}=0.005\%$, and simultaneously, a low assignment error $\epsilon_{\mathcal{F},\textrm{leak}}=0.87\%$.
\section{\label{appendix:measurement_induced_2_0_relaxation} Analysis of $|2\rangle$$\rightarrow$$|0\rangle$ measurement-induced state transition}
\begin{figure}
\includegraphics[width=1\linewidth]{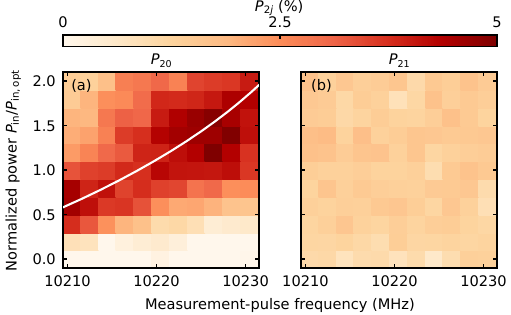}
\caption{\label{fig:measurement_induced_2_0_relaxation_appendix}  Characterization of the $|2\rangle$$\rightarrow$$|0\rangle$ transition. (a),(b)~Extracted seepage probabilities $P_{20}$ and $P_{21}$, respectively, versus the carrier frequency and drive power of the repeated measurement pulse in the MIST benchmarking experiment. The drive power $P_\mathrm{in}$ is expressed relative to the optimal power $P_\mathrm{in,opt}$ used for two-state assignment. The white curve in (a)~indicates the predicted resonance condition for a four-wave scattering process in which a drive photon combines with two photons from the $|2\rangle$$\rightarrow$$|0\rangle$ transition to excite a spurious mode at \qty{26.3}{\giga\hertz} (see Appendix~\ref{appendix:measurement_induced_2_0_relaxation} text).}
\end{figure}

In this Appendix, we analyze the measurement-induced $|2\rangle$$\rightarrow$$|0\rangle$ seepage transition visible in the leakage-sensitive measurement signal in Fig.~\ref{fig:Leakage_sensitive_measurement}. We apply the MIST benchmarking pulse sequence described in the main text, with the qubit initialized in $|2\rangle$, and sweep the carrier frequency and drive power of the repeated measurement. Figures~\ref{fig:measurement_induced_2_0_relaxation_appendix}(a) and (b) show the fitted per-measurement transition probabilities $P_{20}$ and $P_{21}$, respectively, against carrier frequency and drive power. The probability $P_{21}$ shows no correlation with the readout frequency or drive power, with a mean rate $P_{21}=\num{1.4e-2}$ corresponding to a relaxation time $T_1^{2\rightarrow1}=t_\mathrm{meas}/P_{21}=\SI{6.9}{\micro\second}$ with $t_\mathrm{meas}=\SI{96}{\nano\second}$, consistent with background $|2\rangle$$\rightarrow$$|1\rangle$ relaxation. The probability $P_{20}$ depends strongly on both carrier frequency and drive power. At zero drive power, the mean fitted rate is negligible, $P_{20}^{\mathrm{off}}=\num{0(1)e-3}$, as expected since the direct $|2\rangle$$\rightarrow$$|0\rangle$ transition is forbidden to leading order by the selection rules of the transmon~\cite{koch2007charge}. At nonzero drive power, $P_{20}$ increases, with the maximum occurring at a frequency-dependent power.
\\
\indent We hypothesize that the $|2\rangle$$\rightarrow$$|0\rangle$ seepage results from a four-wave scattering process in which a single drive photon combines with the two photons released by the $|2\rangle$$\rightarrow$$|0\rangle$ transition to excite a spurious mode~\cite{connolly2025full, dai_characterization_2026}. The resonance condition for this process is
\begin{equation}
    \omega_\mathrm{d} + \tilde{\omega}_{02} = \omega_\mathrm{s},
    \label{eq:four_wave_resonance_condition}
\end{equation}
where $\omega_\mathrm{d}$ is the drive frequency, $\omega_\mathrm{s}$ is the spurious mode frequency, and $\tilde{\omega}_{02}$ is the dressed $|0\rangle$--$|2\rangle$ energy difference. We write $\tilde{\omega}_{02}=\omega_{02} + \Delta_{02}(\omega_\mathrm{d}, P)$, where $\Delta_{02}(\omega_\mathrm{d}, P)$ is the Stark shift induced by the measurement drive at power $P$. We predict $\Delta_{02}$ using the input--output model. Equation~\eqref{eq:readout_photon_number_vs_drive_frequency} gives the steady-state readout photon number $n_\mathrm{r}^2(\omega_\mathrm{d}, P)$ with the qubit in $|2\rangle$. The Stark shift of the $|0\rangle$--$|2\rangle$ transition is then given by
\begin{equation}
    \Delta_{02}(\omega_\mathrm{d}, P) = \left(\chi_2 - \chi_0\right)n_\mathrm{r}^2(\omega_\mathrm{d}, P),
    \label{eq:stark_shift_0_2}
\end{equation}
where $\chi_j$ is the cavity pull for qubit state $|j\rangle$, which we solve for numerically (see Appendix~\ref{appendix:leakage_sensitive_measurement} for details). For each drive frequency, we solve Eqs.~\eqref{eq:four_wave_resonance_condition} and \eqref{eq:stark_shift_0_2} for the drive power $P$ satisfying the resonance condition, with the spurious mode frequency $\omega_\mathrm{s}$ as the sole fit parameter. The white curve in Fig.~\ref{fig:measurement_induced_2_0_relaxation_appendix}(a) shows the predicted resonance condition for $\omega_s/2\pi=\SI{26.3}{\giga\hertz}$, in good correspondence with the measured power and frequency dependence of the $|2\rangle$$\rightarrow$$|0\rangle$ transition. This measurement-induced transition is therefore consistent with a four-wave scattering process that relaxes the qubit from $|2\rangle$ to $|0\rangle$ and excites a spurious mode at \SI{26.3}{\giga\hertz}.
\section{\label{appendix:MIST} MIST benchmarking}
In this Appendix, we describe the model used to extract the per-measurement measurement-induced-state-transition (MIST) rates from the MIST benchmarking experiment. We then analyze the power dependence of the MIST rates. 
\subsection{\label{appendix:G1subsec} Three-class transition model}
\begin{figure}
\includegraphics[width=1\linewidth]{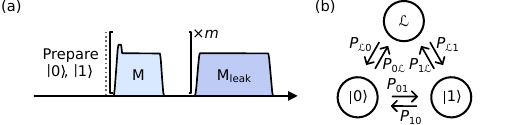}
\caption{\label{fig:MIST_transition_model}  MIST benchmarking. (a) General pulse sequence. The terms $\mathrm{M}$ and $\mathrm{M}_\mathrm{leak}$ denote the measurement being benckmarked and the leakage-sensitive measurement. (b) Three-class transition model. The qubit belongs to one of the three classes $|0\rangle$, $|1\rangle$, and $\mathcal{L}$, where class $\mathcal{L}$ represents all possible leakage states. At each measurement pulse in the repeated measurement block, the qubit has a fixed probability $P_{ij}$ to transition from class $i$ to class $j$.}
\end{figure}

The general pulse sequence for the MIST benchmarking experiment is shown in Fig.~\ref{fig:MIST_transition_model}(a). Figure~\ref{fig:MIST_transition_model}(b) shows the three-class model used to extract the MIST rates. The qubit belongs to one of the three classes $|0\rangle$, $|1\rangle$ or $\mathcal{L}$. At every measurement in the repeated measurement block, the qubit has a fixed probability $P_{ij}$ to transition from class $i$ to class $j$. This model is an approximation for the seepage rates, $P_{\mathcal{L}1}$ and $P_{\mathcal{L}0}$, since the qubit may transition between different leakage states within class $\mathcal{L}$, with each having a different seepage rate back to the computational basis. However, for the qubit initially prepared in $|0\rangle$ or $|1\rangle$, the leakage populations for low repetition number $m$ are to leading order determined solely by the leakage rates $P_{0\mathcal{L}}$ and $P_{1\mathcal{L}}$ and are insensitive to the seepage rates. Since the three-class model reproduces the measured leakage populations well, including at low repetition number $m$, the fitted leakage rates are well constrained.
\\
\indent The class populations $\mathbf{p} = (p_0, p_1, p_{\mathcal{L}})^{\mathrm{T}}$
evolve under the discrete-time Markov chain
\begin{equation}
\mathbf{p}_{m+1} = \left(\mathds{1} + G\right)\mathbf{p}_m \mathrm{,}
\label{eq:MIST_markov}
\end{equation}
\begin{equation}
G =
\begin{pmatrix}
-(P_{01}+P_{0\mathcal{L}}) & P_{10} & P_{\mathcal{L}0}\\
P_{01} & -(P_{10}+P_{1\mathcal{L}}) & P_{\mathcal{L}1}\\
P_{0\mathcal{L}} & P_{1\mathcal{L}} & -(P_{\mathcal{L}0}+P_{\mathcal{L}1})
\end{pmatrix},
\label{eq:MIST_markov_matrix}
\end{equation}
where $m$ is the measurement index and the columns of $G$ sum to zero,
conserving probability. For $P_{ij} \ll 1$, the populations are well approximated by the solution of the rate equation
\begin{equation}
\frac{d\mathbf{p}}{dm} = G\,\mathbf{p} \mathrm{,}
\label{eq:MIST_rate_equation}
\end{equation}
with $m$ treated as a continuous variable. The transition probabilities
$P_{ij}$ are extracted from a simultaneous least-squares fit of
Eq.~\eqref{eq:MIST_rate_equation} to the measured populations for the
qubit prepared in $|0\rangle$ and $|1\rangle$.
\subsection{\label{appendix:G2subsec} MIST power dependence}
\indent We repeated the MIST benchmarking experiment while varying the power
$P_\mathrm{in}$ of the measurement pulses in the repeated measurement block, in
order to investigate the power dependence of the MIST rates. Throughout this
section, $P_\mathrm{in}$ is expressed relative to the optimal power
$P_\mathrm{in,opt}$ used for two-state assignment.
\begin{figure*}
\includegraphics[width=1\linewidth]{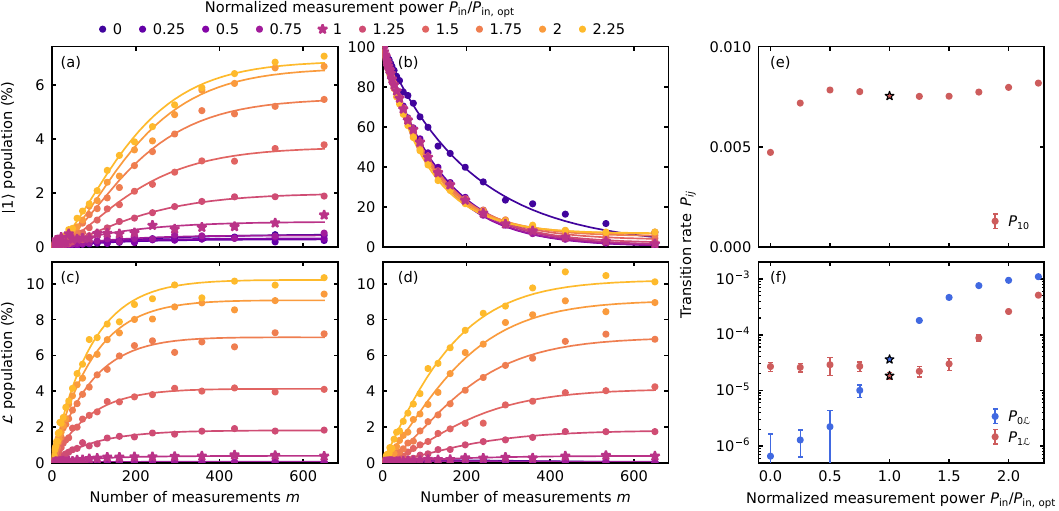}
\caption{\label{fig:MIST_vs_power_appendix} MIST rates as a function of the measurement power. (a),(b)~Final $|1\rangle$-class populations, for the qubit prepared in $|0\rangle$ and $|1\rangle$, respectively, at different measurement powers. (c),(d)~Final $\mathcal{L}$-class populations, for the qubit prepared in $|0\rangle$ and $|1\rangle$, respectively, at different measurement powers. (e)~Measurement-induced relaxation rate $P_{10}$ against measurement power. (f)~Measurement-induced leakage rates $P_{0\mathcal{L}}$ and $P_{1\mathcal{L}}$ against measurement power. In all panels, the measurement power $P_\mathrm{in}$ is expressed relative to the optimal power $P_\mathrm{in,opt}$ for two-state assignment. Measurements performed at the optimal power are indicated by star markers.}
\end{figure*}
\\
\indent Figures~\ref{fig:MIST_vs_power_appendix}(a) and (b) show the final
$|1\rangle$-class populations for the qubit prepared in $|0\rangle$ and
$|1\rangle$, respectively, and Figs.~\ref{fig:MIST_vs_power_appendix}(c) and
(d) show the final $\mathcal{L}$-class populations. As the measurement power increases, the leakage rates increase, and the
steady-state leakage population grows from approximately $0.4\%$ at
$P_\mathrm{in,opt}$ to $10\%$ at $2.25 \times P_\mathrm{in,opt}$. Figure~\ref{fig:MIST_vs_power_appendix}(e) shows the transition probability
$P_{10}$ against the measurement power. With the measurement off (zero power),
we obtain $P_{10}^{\mathrm{off}} = 4.73\times10^{-3}$, which corresponds to a relaxation time $T_1 = t_\mathrm{meas}/P_{10}^{\mathrm{off}} =
\SI{20.3}{\micro\second}$, taking $t_\mathrm{meas} = \SI{96}{\nano\second}$,
consistent with the relaxation time $T_1=\SI{19}{\micro\second}$ determined in a standard $T_1$ relaxation experiment. As the power increases, $P_{10}$ rises and then plateaus.
\\
\indent Figure~\ref{fig:MIST_vs_power_appendix}(f) shows the leakage probabilities
$P_{0\mathcal{L}}$ and $P_{1\mathcal{L}}$ against the measurement power. The probability $P_{0\mathcal{L}}$ is negligible at zero power and increases with power, exceeding $P_{1\mathcal{L}}$ near $P_\mathrm{in,opt}$. The probability $P_{1\mathcal{L}}$ is approximately constant for powers below $\sim\!1.5\,P_\mathrm{in,opt}$, attributed to background $|1\rangle$$\rightarrow$$|2\rangle$ leakage, above which it increases with power. The small dip in $P_{1\mathcal{L}}$ at $P_\mathrm{in,opt}$ may be a fitting artifact since the deviation is comparable to the fit uncertainty. At twice the optimal power, leakage from both $|0\rangle$ and $|1\rangle$ rises by more
than an order of magnitude: $P_{0\mathcal{L}}$ increases from
$\num{3.6(3)e-5}$ to $\num{9(1)e-4}$ and $P_{1\mathcal{L}}$ from $\num{1.8(3)e-5}$
to $\num{2.6(1)e-4}$.

\section{\label{appendix:multiphoton_simulations} Multiphoton resonances}
In this appendix, we analyze the hybridization parameter $\Theta_j$ and the quasifrequency collisions as a function of the transmon gate charge $n_\mathrm{g}$. We then describe how the leakage probabilities for a given quasifrequency collision are calculated using the Landau-Zener formula. We show that, due to the low photon numbers driven by the two-state measurement pulse, the leakage probability due to multiphoton resonances is small regardless of the gate charge.
\begin{figure}
\includegraphics[width=1\linewidth]{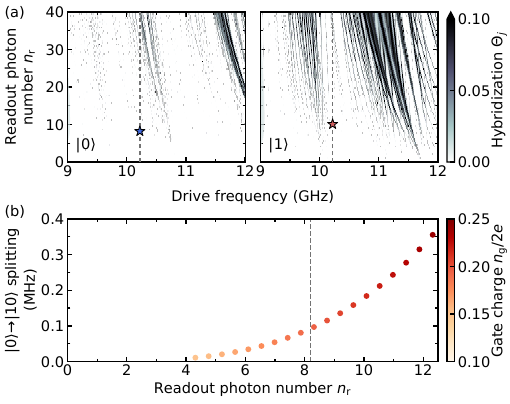}
\caption{\label{fig:multiphoton_resonances_appendix} Multiphoton resonance simulations. (a)~Floquet simulations of the hybridization parameter $\Theta_j$ for the qubit in the $|0\rangle$~(left) and $|1\rangle$~(right) states. The results for ten gate charge values evenly spaced between $n_\mathrm{g}/2e=0$ and $n_\mathrm{g}/2e=0.45$ are overlaid. The drive frequency and maximum photon numbers for the two-state measurement in the experiments are indicated by the star markers. (b)~Simulated avoided-crossing splitting $\Delta_\varepsilon$ versus readout photon number for the $|0\rangle \rightarrow |10\rangle$ quasifrequency collision, as a function of the gate charge. Collisions with splittings below $10~\mathrm{kHz}$ are omitted. The gray dashed line indicates the maximum photon number driven by the two-state measurement pulse.}
\end{figure}
\subsection{\label{appendix:H1subsec} Multiphoton resonances versus gate charge}
Figure~\ref{fig:multiphoton_resonances_appendix}(a) shows the hybridization parameter $\Theta_j$ overlaid for ten evenly-spaced gate charge values between $n_\mathrm{g}/2e=0$ and $n_\mathrm{g}/2e=0.45$. The multiphoton resonances are weak at the maximum photon number induced by the two-state measurement pulse, irrespective of the gate charge.
\\
\indent Figure~\ref{fig:multiphoton_resonances_appendix}(b) shows the dominant $|0\rangle$$\rightarrow$$|10\rangle$ quasifrequency-collision splitting $\Delta_\varepsilon$ against the readout photon number at which the collision occurs~(which varies with the gate charge $n_\mathrm{g}$), for a readout drive at frequency $\omega_\mathrm{d}/2\pi = 10220.5~\mathrm{MHz}$. The avoided-crossing splitting increases with increasing readout photon number; however, the splitting is weak with $\Delta_\varepsilon/2\pi < 0.1$~MHz for readout photon numbers up to the maximum induced by the two-state measurement pulse, $n_{\mathrm{r},\mathrm{max}}^0=8.2$, given the qubit in $|0\rangle$.
\subsection{\label{appendix:H2subsec} Landau-Zener leakage probability}
\indent To predict the adiabatic leakage probability $P_\mathrm{L}$ for a quasifrequency collision traversal we use the Landau-Zener formula ${P_\mathrm{L} = 1-\mathrm{exp}\left(-\pi \Delta_\varepsilon^2 /2\upsilon \right)}$,  which applies well to Floquet avoided crossings~\cite{ikeda2022floquet, dumas_measurement-induced_2024} and has been experimentally verified in the context of multiphoton-resonance induced leakage in transmon qubits~\cite{wang2026probing}. The term $\upsilon$ is the crossing speed. Denoting the lower branch quasifrequency as $\varepsilon_\mathrm{l}$ and the upper branch quasifrequency as $\varepsilon_\mathrm{u}$, the crossing speed $\upsilon$ takes the form~\cite{ikeda2022floquet, dumas_measurement-induced_2024}
\begin{equation}
    \upsilon = \left(2\Delta_{\varepsilon}\right)^{1/2} \left|\frac{d^2 \varepsilon_\mathrm{l}}{d n_\mathrm{r} ^2}\right|^{1/2}_{n_{\mathrm{r}, \mathrm{cross}}} \dot{n}_\mathrm{r}(t_\mathrm{cross}) \, \mathrm{,}
\label{eq:LZ_crossing_speed_appendix}
\end{equation}
where this assumes that the second derivatives for $\varepsilon_\mathrm{l}$ and $\varepsilon_\mathrm{u}$ are equal at the crossing point. The terms $n_{\mathrm{r}, \mathrm{cross}}$ and $t_\mathrm{cross}$ are the readout photon number and the time at the crossing, respectively, and the sweep rate $\dot{n}_\mathrm{r}(\tau)$ is defined,
\begin{equation}
\dot{n}_\mathrm{r}(\tau) \equiv \left|\frac{d n_r(t)}{dt}\right|_{t=\tau} \, \mathrm{.}
\label{eq:leakage_susceptibility}
\end{equation}
The sweep rate carries all the time dependence, with the remaining terms in Eq.~\eqref{eq:LZ_crossing_speed_appendix} fully determined by the static Floquet simulations. To leading order, the adiabatic leakage probability is proportional to the inverse sweep rate, $P_\mathrm{L}\propto 1/\dot{n}_\mathrm{r}(\tau)$. During the ring-up and ring-down, the sweep rate is proportional to the effective readout decay rate, $\dot{n}_\mathrm{r} \propto \kappa_{\mathrm{eff}}$. Thus, due to the large effective decay rate of the readout resonator, the leakage probability is reduced during both the ring-up and the ring-down quasifrequency collision traversals. We treat both traversals as independent and sum their contributions to find the total leakage probability. We solve for the sweep rate numerically using the input--output model. For 100 evenly sampled gate charge values between $n_\mathrm{g}/2e=0$ and $n_\mathrm{g}/2e=0.495$, we find that the dominant $|0\rangle$$\rightarrow$$|10\rangle$ transition has mean leakage probability $P_\mathrm{L}=\num{2e-8}$ and maximum $P_\mathrm{L}=\num{1e-6}$. Repeating the analysis for the dominant $|1\rangle$$\rightarrow$$|9\rangle$ transition (not shown), the mean leakage probability is $P_\mathrm{L}=\num{1e-7}$ while the maximum is $P_\mathrm{L}=\num{2e-6}$. Multiphoton resonances are therefore predicted to contribute negligible leakage during the two-state measurement pulse for the qubit characterized in this work.

\bibliography{bib} 
\end{document}